\documentclass[11pt]{article}
\usepackage{subfig}
\usepackage[mathscr]{euscript}
\usepackage{color}
\usepackage{array}
\usepackage[english]{babel}
\usepackage{booktabs}
\usepackage{caption}
\usepackage[numbers]{natbib}
\renewcommand{\cite}{\citep}

\usepackage[margin=1in]{geometry}

\usepackage{subfig}

\newcommand{\mc}{\circ} 
\newcommand{\ec}[2]{#1 \ast_E #2} 
\newcommand{\rr}[1]{\mathsf{#1}} 
\newcommand{\dontprintsemicolon}{\DontPrintSemicolon}
\usepackage[noend,ruled,linesnumbered,algosection]{algorithm2e}
\SetKw{Break}{break}
\SetKw{Abort}{abort}
\SetKw{Continue}{continue}

\usepackage{amsmath, amssymb}
\usepackage{mathtools}
\usepackage{tikz}
\usetikzlibrary{shapes,positioning,arrows,calc}

\newcommand{\ignore}[1]{}

\usetikzlibrary{matrix,arrows}

\usepackage{graphvizObabel}

\date{}
\begin{document}

\title{Inferring Chemical Reaction Patterns Using\\ Rule Composition in Graph Grammars}
\author{Jakob L. Andersen,$^{1}$ Christoph
  Flamm,$^{2}$ Daniel Merkle,$^1$\footnote{to whom correspondence should be
    addressed}\,\, and Peter~F.~Stadler\,$^{2-7}$\\[0.5cm]
   $^1$ Department for Mathematics and Computer Science\\ 
   University of Southern Denmark, Denmark\\[0.2cm]
  $^2$ Institute for Theoretical Chemistry\\University of Vienna, Austria.\\[0.2cm]
   $^3$ Bioinformatics Group, Department of Computer Science\\
   and Interdisciplinary Center for
   \phantom{$^3$}Bioinformatics\\University of Leipzig, Germany.\\[0.2cm]
   $^4$ Max Planck Institute for Mathematics in the Sciences, Leipzig, Germany.\\[0.2cm]
   $^5$ Fraunhofer Institute for Cell Therapy and Immunology, Leipzig, Germany.\\[0.2cm]
   $^6$ Center for non-coding RNA in Technology and Health\\
   University of Copenhagen, Denmark.\\[0.2cm]
   $^7$ Santa Fe Institute, 1399 Hyde Park Rd, Santa Fe, NM 87501, USA
}%



\maketitle

\begin{abstract}
  Modeling molecules as undirected graphs and chemical reactions as
  graph rewriting operations is a natural and convenient approach to
  modeling chemistry. Graph grammar rules are most naturally employed
  to model elementary reactions like merging, splitting, and
  isomerisation of molecules. It is often convenient, in particular in
  the analysis of larger systems, to summarize several subsequent
  reactions into a single composite chemical reaction. We use a
  generic approach for composing graph grammar rules to define a
  chemically useful rule compositions. We iteratively apply these rule
  compositions to elementary transformations in order to automatically
  infer complex transformation patterns. This is useful for instance
  to understand the net effect of complex catalytic cycles such as the
  Formose reaction. The automatically inferred graph grammar rule is a
  generic representative that also covers the overall reaction pattern
  of the Formose cycle, namely two carbonyl groups that can react with
  a bound glycolaldehyde to a second glycolaldehyde. Rule composition
  also can be used to study polymerization reactions as well as more
  complicated iterative reaction schemes. Terpenes and the
  polyketides, for instance, form two naturally occurring classes of
  compounds of utmost pharmaceutical interest that can be understood
  as ``generalized polymers'' consisting of five-carbon (isoprene) and
  two-carbon units, respectively.

\end{abstract}

\section{Introduction}
Directed hypergraphs \cite{Klamt:2009} are a suitable topological
representation of (bio)che\-mi\-cal reaction networks where
(catalytic) reactions are hyperedges connecting substrate nodes to
product nodes. Such networks require an underlying Artificial
Chemistry \cite{Dittrich:01} that describes how molecules and
reactions are modeled. If molecules are treated as edge and vertex
labeled graphs, where the vertex labels correspond to atom types and
the edge labels denote bond types, then structural change of molecules
during chemical reactions can be modeled as graph rewrite
\cite{Benkoe:2003}. In contrast to many other Artificial Chemistries
this approach allows for respecting fundamental rules of chemical
transformations like mass conservation, atomic types, and cyclic
shifts of electron pairs in reactions. In general, a graph rewrite
(rule) transforms a set of substrate graphs into a set of product
graphs. Hence the graph rewrite formalism allows not only to delimit
an entire chemical universe in an abstract but compact form but also
provides a methodology for its explicit construction.

Most methods for the analysis of this network structure are directed
towards this graph (or hypergraph) structure
\cite{Aittokallio:2006,Klamt:2009}, which is described by the
stoichiometric matrix $\mathbf{S}$ of the chemical system. Since
$\mathbf{S}$ is essentially the incidence matrix of the directed
hypergraph, algebraic approaches such as Metabolic Flux Analysis and Flux
Balance Analysis \cite{Orth:10} have a natural interpretation in terms of
the hypergraph. Indeed typical results are sets of possibly weighted
reactions (i.e., hyperedges) such as elementary flux modes
\cite{Schuster:00}, extreme pathways \cite{Price:03}, minimal metabolic
behaviors \cite{Larhlimi:09} or a collection of reactions that maximize the
production of a desired product in metabolic engineering. The net reaction
of a given pathway is simply the linear combination of the participating
hyperedges.

In the setting of generative models of chemistry, each concrete reaction is
not only associated with its stoichiometry but also with the transformation
rule operating on the molecules that are involved in a particular
reaction. Importantly, these rules are formulated in terms of
\emph{reaction mechanisms} that readily generalize to large sets of
structurally related molecules. It is thus of interest to derive not only
the stoichiometric net reaction of a pathway but also the corresponding
``effective transformation rule''. Instead of attempting to address this
issue \emph{a posteriori}, we focus here on the possibility of composing
the elementary rules of chemical transformations to new effective rules
that encapsulate entire pathways.
  
The motivation comes from the observation that string grammars are
meaningfully characterized and understood by investigating the
transformation rules. Consider, as a trivial example, the context-free
grammar $\mathscr{G}$ with the starting symbol $S$ and the rules $S
\rightarrow aS'a$, $S'\rightarrow aS'a\: |\: B$ and $B\rightarrow
\epsilon\: |\: bB$. Inspecting this grammar we see that we can summarize
the effect of the productions as $B\rightarrow b^k$, $k\ge0$, and
$S\rightarrow a^n B a^n$, $n\ge 1$. The language generated from
$\mathscr{G}$ is thus $\{a^nb^*a^n| n\geq 1\}$. Here we explore whether a
similar reasoning, namely the systematic combination of transformation
rules, can help to characterize the language of molecules that is generated
by a particular graph rewriting chemistry. Similar to the example from term
rewriting above, we should at the very least be able to recognize the
regularities in polymerization reactions. We shall see below, however, that
the rule based approach holds much higher promises.

In this contribution we address two issues: First we establish the formal
conditions under which chemical transformation rules \emph{can} be
meaningfully composed. To this end, we introduce in
section~\ref{sec:gg-and-rc} rule composition within the framework of
concurrency theory. We then discuss the specific restrictions that apply to
chemical systems, leading to the constructive approach to inferring
composed rules in section~\ref{sec:cons}.

The basic computational task we envision starts from an unordered set
$\mathscr{R}$ of reactions such as those forming a particular metabolic
reaction pathway. To derive the effective transformation rule describing
the pathway we need to find the correct ordering $\pi$ in which the
transformation rules $p_i$, underlying the individual chemical reactions
$\rho_i$, have to be composed. We illustrate this approach in some detail
using the Formose reaction as an example in Section~\ref{sec:res}.

\section{Graph Grammars and Rule Composition}
\label{sec:gg-and-rc}

Graph grammars, or graph rewriting systems, are proper generalizations of
term rewrite systems. A wide variety of formal frameworks have been
explored, including several different algebraic ones rooted in category
theory. As a model of chemical transformations the so-called \emph{double
  pushout} (DPO) formulation appears to be best suited. We refer to
\cite{Ehrig:06} for the comprehensive treatise. In the following sections
we first outline the basic setup and then introduce full and partial rule
composition.

\subsection{Double Pushout and Concurrency}

The DPO formulation of graph transformations considers transformation rules
of form $p=(L\xleftarrow{l}K\xrightarrow{r}R)$ where $L$, $R$, and $K$ are
called the left graph, right graph, and context graph, respectively. The
maps $l$ and $r$ are graph morphisms. The rule $p$ transforms $G$ to $H$,
in symbols $G\xRightarrow{p,m} H$ if there is a pushout graph $D$ and a
``matching morphism'' $m:L\to G$ such that following diagram is valid:
\begin{equation}
\begin{minipage}[c]{0.8\columnwidth}
\begin{center}
\vspace*{-1.0em}
\begin{tikzpicture}[description/.style={fill=white,inner sep=2pt}]
\matrix (n) [matrix of math nodes, row sep=2em,
column sep=2.5em, text height=1.5ex, text depth=0.25ex]
{ L & & K & & R \\
  G & & D & & H \\ };
\path[->,font=\scriptsize]
(n-1-3) edge node[description] {$ l $} (n-1-1);
\path[->,font=\scriptsize]
(n-1-3) edge node[description] {$ r $} (n-1-5);
\path[->,font=\scriptsize]
(n-2-3) edge node[description] {$\strut \rho $} (n-2-1);
\path[->,font=\scriptsize]
(n-2-3) edge node[description] {$\strut \lambda $} (n-2-5);
\path[->,font=\scriptsize]
(n-1-1) edge node[description] {$\phantom{k}m\phantom{k}$} (n-2-1);
\path[->,font=\scriptsize]
(n-1-3) edge node[description] {$ k $} (n-2-3);
\path[->,font=\scriptsize]
(n-1-5) edge node[description] {$\phantom{k}n\phantom{k}$} (n-2-5);
\end{tikzpicture}
\vspace*{-0.7em}
\end{center}
\end{minipage}
\label{eq:rule}
\end{equation}
The existence of $D$ is equivalent to the so-called \emph{gluing
  condition}, which determines whether the rule $p$ is applicable to a
match in $G$. In the following we will also write $G\xRightarrow{p} H$ and
$G\Rightarrow H$ for derivations, if the specific match or transformation
rule is unimportant or clear from the context.

Concurrency theory provides a canonical framework for the composition of
two graph transformations. Given two rules
$p_i=(L_i\xleftarrow{l_i}K_i\xrightarrow{r_i}R_i)$, $i=1,2$, a composition
$(L\xleftarrow{q_l} K\xrightarrow{q_r} R) = \ec{p_1}{p_2}$ can be defined
whenever a dependency graph $E$ exists so that in the following diagram:
\begin{equation}
\begin{minipage}[c]{0.9\columnwidth}
\begin{center}
\scriptsize
\vspace*{-1.0em}
\begin{tikzpicture}[description/.style={fill=white,inner sep=2pt}]
\matrix (m) [matrix of math nodes, row sep=0.3em,
column sep=0.8em, text height=1.5ex, text depth=0.25ex]
{ L_1 & & K_1 & & R_1 &   & L_2 & & K_2 & & R_2 \\
&\phantom{(0)} &     &(1)&   &   &     &(2)&   & &     \\
  L   & & C_1 & &     & E &     & & C_2 & &  R  \\
      & &     & &     &(3)&     & &     & &     \\
      & &     & &     & K &     & &     & \phantom{(0)}&     \\ };
\path[->,font=\scriptsize]
(m-1-3) edge node[description] {$ l_1 $} (m-1-1);
\path[->,font=\scriptsize]
(m-1-3) edge node[description] {$ r_1 $} (m-1-5);
\path[->,font=\scriptsize]
(m-3-3) edge node[description] {$ s_1 $} (m-3-1);
\path[->,font=\scriptsize]
(m-3-3) edge node[description] {$ t_1 $} (m-3-6);
\path[->,font=\scriptsize]
(m-1-1) edge node[description] {$ u_l $} (m-3-1);
\path[->,font=\scriptsize]
(m-1-3) edge node[description] {$ v_1 $}   (m-3-3);
\path[->,font=\scriptsize]
(m-1-5) edge node[description] {$ e_1 $} (m-3-6);
\path[->,font=\scriptsize]
(m-1-9) edge node[description]       {$ l_2 $} (m-1-7);
\path[->,font=\scriptsize]
(m-1-9) edge node[description] {$ r_2 $} (m-1-11);
\path[->,font=\scriptsize]
(m-3-9) edge node[description] {$ s_2 $} (m-3-6);
\path[->,font=\scriptsize]
(m-3-9) edge node[description] {$ t_2 $} (m-3-11);
\path[->,font=\scriptsize]
(m-1-7) edge node[description] {$ e_2 $} (m-3-6);
\path[->,font=\scriptsize]
(m-1-9) edge node[description] {$ v_2 $} (m-3-9);
\path[->,font=\scriptsize]
(m-1-11) edge node[description] {$ u_2 $} (m-3-11);
\path[->,font=\scriptsize]
(m-5-6) edge node[description] {$ w_1 $} (m-3-3);
\path[->,font=\scriptsize]
(m-5-6) edge node[description] {$ w_2 $} (m-3-9);
\end{tikzpicture}
\vspace*{-1.5em}
\end{center}
\end{minipage}
\label{eq:push1}
\end{equation}
the cycles (1) and (2) are pushouts, and (3) is a pullback, see e.g.,
\cite{Golas:10}.  We then have $q_l=s_1\mc w_1$ and $q_r=t_2\mc w_2$.  The
concurrency theorem \cite{Ehrig:91} ensures that for any sequence of
consecutive direct transformations
$G\xRightarrow{p_1,m_1}H\xRightarrow{p_2,m_2} G'$ a graph $E$, a
corresponding $E$-concurrent rule $\ec{p_1}{p_2}$, and a morphism $m$ can be
found such that $G \xRightarrow{\ec{p_1}{p_2},m} G'$.

In order to use graph transformation as a model for chemical reactions
additional conditions must be enforced. Most importantly, atoms are neither
created, nor destroyed, nor transformed to other types. Thus only graph
morphisms whose restriction to the vertex sets are bijective are valid in
our context. In particular, the matching morphism $m$ always corresponds 
to a subgraph isomorphism in our context. The context graph $K$ thus 
is (isomorphic to) a subgraph of both $L$ and $R$, describing the part of 
$L$ that remains unchanged in $R$. Conservation of atoms means that the 
vertex sets of $L$, $K$, and $R$ are linked by bijections known as the 
atom-mapping. When the atom mapping is clear, thus, we do not need 
to represent the context explicitly.

It is important to note that the existence of the matching morphism $m:L\to
G$ alone is not sufficient to guarantee the applicability of the
transformation. In our context, we require in addition that the
transformation rule does not attempt to introduce an edge in $R$ that has
been present already before the transformation is applied. Formally, the
\emph{gluing condition} requires that $(l(x),l(y))\notin L$ and
$(r(x),r(y))\in R$ implies $(m(l(x)),m(r(y)))\notin G$.

\subsection{Full Rule Composition}

\begin{figure}
  \scriptsize
  \begin{center} 
    \begin{tikzpicture}[
      node distance=0.7cm,
      large/.style={draw,shape=ellipse,minimum height=18pt, 
        minimum width=50pt},
      small/.style={draw,shape=ellipse,minimum height=7pt, 
        minimum width=8pt}, 
      normal/.style={->,>=triangle 45}, 
      ]
      \draw node[large](L1){$L_1$} node[large](R1)[right=of L1]{}
      node[left] at (R1){$R_1$} node[small,right](L2) at (R1){$L_2$}
      node[small](R2)[right=of L2]{$R_2$} ; \draw[normal](L1) to
      node[auto]{$p_1$} (R1); \draw[normal](L2) to node[auto]{$p_2$}
      (R2);
    \end{tikzpicture}
    ~\\[0.1cm]$\Longrightarrow$\\[0.3cm]
    \begin{tikzpicture}[
      node distance=0.7cm,
      large/.style={draw,shape=ellipse,minimum height=18pt, 
        minimum width=50pt},
      small/.style={draw,shape=ellipse,thin,dashed,minimum height=7pt, 
        minimum width=8pt}, 
      normal/.style={->,>=triangle 45}, 
      ]
      \draw node[large](L3){$L_1$} node[large](R3)[right=of L3]{}
      node[left] at (R3){$R_3$} node[small,right](R2) at (R3){$R_2$} ;
      \draw[normal](L3) to node[auto]{$p_3$} (R3);
    \end{tikzpicture}
\end{center}
\caption[Full rule composition]{Full composition of two rules requires that
  $L_2$ is (isomorphic to) a subgraph of $R_1$.}
\label{fig:compsimp}
\end{figure}
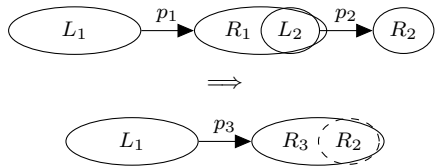

In the following we will be concerned only with special, chemically
motivated, types of rule compositions. In the simplest case the dependency
graph $E$ is isomorphic to $R_1$, later we will also consider a more
general setting in which $E$ is isomorphic to the disjoint union of $R_1$
and some connected components of $L_2$. For the ease of notation from now
on we only refer to a rule composition, and not to a composition of
morphisms as in Section \ref{sec:gg-and-rc}, i.e., $\ec{p_1}{p_2}$ will be
denoted as $p_2 \mc p_1$ (note the order of the arguments changes). If $E
\cong R_1$, then $L_2\cong e_2(L_2)$ is a subgraph of $R_1$.  Omitting the
explicit references to the subgraph matching morphism $e_2$ we can simply
view $L_2$ as subgraph of $R_1$ as illustrated in
Figure~\ref{fig:compsimp}.

The rule composition thus amounts to a rewriting
$R_1\xRightarrow{p_2,e_2}R$, while the left side $L_1$ is preserved. We
will use the notation $p_2\circ p_1$ and $G'\xRightarrow{p_2 \circ p_1} G'$
for this restricted type of rule composition, and call it \emph{full}
composition as the complete left side of $p_2$ is a subgraph of $R_1$.
Note that $L_2$ may fit into $R_1$ in more than one way so that there may
be more than one composite rule. Formally, the alternative compositions are
distinguished by different matching morphisms $e_2$ in the diagram
(\ref{eq:push1}); we will return to this point below.

\subsection{Partial Rule Composition}

\begin{figure}[b]
  \small
  \begin{center} 
    \begin{tikzpicture}[
      node distance=0.6cm,
      large/.style={draw,shape=ellipse,minimum height=28pt, 
        minimum width=46pt},
      small/.style={draw,shape=ellipse,minimum height=7pt, 
        minimum width=12pt},
      normal/.style={->,>=triangle 45},
      ]
      \draw	
      node[large](L1){$\phantom{^1_2}L_1\phantom{^1_2}$}
      node[large](R1)[right=of L1]{} node[left] at (R1){$\phantom{^1_2}R_1$}
      node[small,left](L21) at (R1.east){$L_2^1$}
      node(L2)[below=0.1cm of L21]{}
      node[](Merge2)[right=of L2]{} node[above] at (Merge2){$p_2$}
      node[small](L22)[below=0.1cm of L2]{$L_2^2$}
      node[small](R2)[right=of Merge2]{$R_2$};
      \draw[normal](L1) to node[auto]{$p_1$} (R1);
      \draw[normal](L21) to [in=180,out=-43] (R2);
      \draw[normal](L22) to [in=180,out=43] (R2);
    \end{tikzpicture}
    ~\\[0.1cm]$\Longrightarrow$\\[0.3cm]
    \begin{tikzpicture}[
      node distance=0.6cm,
      large/.style={draw,shape=ellipse,minimum height=28pt, 
        minimum width=46pt},
      small/.style={draw,shape=ellipse,minimum height=7pt, 
        minimum width=12pt},
      normal/.style={->,>=triangle 45},
      ]
      \draw	
      node[large](L1){$L_1$}
      node[shape=ellipse,minimum height=10pt, 
      minimum width=17pt,right](L21) at (L1){\phantom{$L^1_2$}}
      node(L2)[below=0.1cm of L21]{}
      node[](Merge2)[right=of L2]{} node[above] at (Merge2){$p_3$}
      node[small](L22)[below=0.1cm of L2]{$L_2^2$}
      node[large](R3)[right=of Merge2]{} node[left] at (R3){$R_3$}
      node[small,left,thin,dashed](R2) at (R3.east){$R_2$};
      \draw[normal](L1) to [in=180,out=-30] (R3);
      \draw[normal](L22) to [in=180,out=43] (R3);
    \end{tikzpicture}
  \end{center}
  \caption[Partial rule composition]{Partial rule composition requires that
    at least one connected component (here $L_2^1$) is isomorphic to
    $R_1$. Additional components of of the second rule may remain unmatched.}
  \label{fig:compnotsimp}
\end{figure}
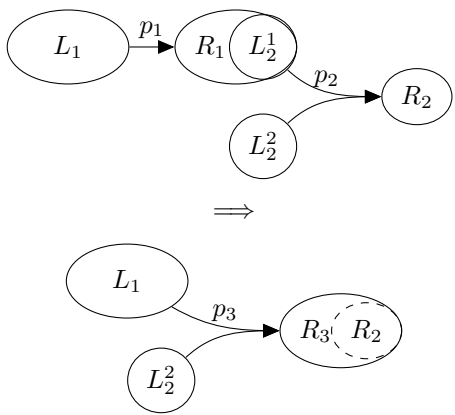

\newcommand\scalingFactor{0.2} \newcommand\fontSize{40}
\newcommand\edgeColour{red} \newcommand\edgePenwidth{3}
\begin{figure*}[t]
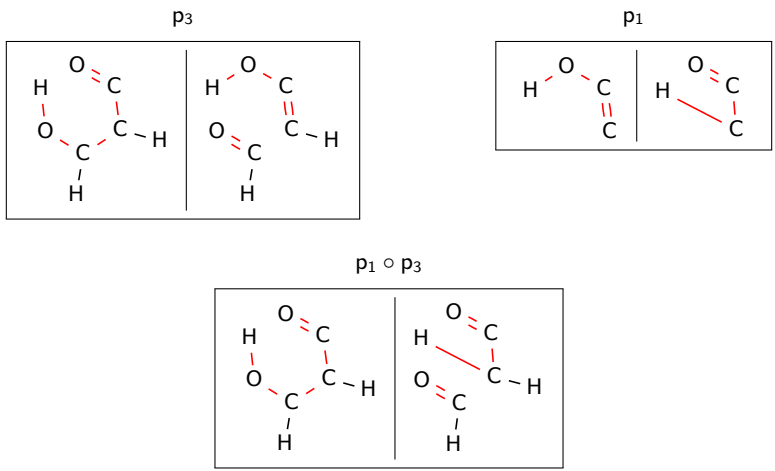

\begin{center}
\captionsetup[subfigure]{position=top,labelformat=empty}
\subfloat[{$\rr{p_3}$}]{
\fbox{
\neatoGraph[trim=1.8cm 0cm 1.2cm 0cm, scale=\scalingFactor]{
node [ shape=plaintext, fontsize=\fontSize ]
edge [ color=\edgeColour, penwidth=\edgePenwidth ]
H
O4 [ label=O ]
C5 [ label=C ]
C0 [ label=C ]
C1 [ label=C ]
O2 [ label=O ]
H -- O4 -- C5 -- C0 -- C1 -- O2 -- C1
H -- O2 [ style=invis ]
edge [ color=black ]
HE1 [ label=H ]
HE2 [ label=H ]
C5 -- HE1
C0 -- HE2
}
\vrule
\neatoGraph[trim=1.2cm 0cm 1.6cm 0cm, scale=\scalingFactor]{
node [ shape=plaintext, fontsize=\fontSize ]
edge [ color=\edgeColour, penwidth=\edgePenwidth ]
H
O4 [ label=O ]
C5 [ label=C ]
C0 [ label=C ]
C1 [ label=C ]
O2 [ label=O ]
C5 -- O4 -- C5
H -- O2 -- C1 -- C0 -- C1
C0 -- C5 [ style=invis ]
O4 -- H [ style=invis ]
edge [ color=black ]
HE1 [ label=H ]
HE2 [ label=H ]
C5 -- HE1
C0 -- HE2
}
}
} \qquad\qquad 
\subfloat[$\rr{p_1}$]{
\fbox{
\neatoGraph[trim=1.8cm 4.7cm 3.8cm 0cm, scale=\scalingFactor]{
node [ shape=plaintext, fontsize=\fontSize ]
edge [ color=\edgeColour, penwidth=\edgePenwidth ]
H
O4 [ label=O, style=invis ]
C5 [ label=C, style=invis ]
C0 [ label=C ]
C1 [ label=C ]
O2 [ label=O ]
C5 -- O4 -- C5 [ style= invis ]
H -- O2 -- C1 -- C0 -- C1
node [ style=invis ]
edge [ style=invis ]
C0 -- C5
O4 -- H
edge [ color=black ]
HE1 [ label=H ]
HE2 [ label=H ]
C5 -- HE1
C0 -- HE2
}
\vrule
\neatoGraph[trim=1.2cm 4.7cm 3.8cm 0cm, scale=\scalingFactor]{
node [ shape=plaintext, fontsize=\fontSize ]
edge [ color=\edgeColour, penwidth=\edgePenwidth ]
H
O4 [ label=O, style=invis ]
C5 [ label=C, style=invis ]
C0 [ label=C ]
C1 [ label=C ]
O2 [ label=O ]
C5 -- O4 -- C5 [ style= invis ]
C1 -- O2 -- C1 -- C0
H -- C0 [ len=2.3 ]
node [ style=invis ]
edge [ style=invis ]
H -- O2 
C0 -- C5 
O4 -- H 
edge [ color=black ]
HE1 [ label=H ]
HE2 [ label=H ]
C5 -- HE1
C0 -- HE2
}
}
}
\qquad\qquad
\subfloat[$\rr{p_1}\circ \rr{p_3}$]{
\fbox{
\neatoGraph[trim=1.8cm 0cm 1.2cm 0cm, scale=\scalingFactor]{
node [ shape=plaintext, fontsize=\fontSize ]
edge [ color=\edgeColour, penwidth=\edgePenwidth ]
H
O4 [ label=O ]
C5 [ label=C ]
C0 [ label=C ]
C1 [ label=C ]
O2 [ label=O ]
H -- O4 -- C5 -- C0 -- C1 -- O2 -- C1
H -- O2 [ style=invis ]
edge [ color=black ]
HE1 [ label=H ]
HE2 [ label=H ]
C5 -- HE1
C0 -- HE2
}
\vrule
\neatoGraph[trim=1.2cm 0cm 1.6cm 0cm, scale=\scalingFactor]{
node [ shape=plaintext, fontsize=\fontSize ]
edge [ color=\edgeColour, penwidth=\edgePenwidth ]
H
O4 [ label=O ]
C5 [ label=C ]
C0 [ label=C ]
C1 [ label=C ]
O2 [ label=O ]
C5 -- O4 -- C5
C1 -- O2 -- C1 -- C0
H -- C0 [ len=2.3 ]
H -- O2 [ style=invis ]
C0 -- C5 [ style=invis ]
O4 -- H [ style=invis ]
edge [ color=black ]
HE1 [ label=H ]
HE2 [ label=H ]
C5 -- HE1
C0 -- HE2
}
}
}
\clearcaptionsetup{subfigure}
\end{center}
\caption[Rule composition]{Composition of two rules from the Formose
  reaction ($\rr{p_1}$ and $\rr{p_3}$); the following rule names will be
  used: $\rr{p_0}$: forward keto-enol tautomerism (which corresponds to the
  reverse of $\rr{p_1}$), $\rr{p_1}$: backward keto-enol tautomerism,
  $\rr{p_2}$: forward aldol addition (which corresponds to the reverse of
  $\rr{p_3}$), and $\rr{p_3}$: backward aldol addition. The atom
    mapping and matching morphism are implicitly given in these drawings by
    corresponding positions of the atoms. The context $K$ thus consists of
    all the atoms as well as the chemical bonds (edges) shown in black in
    both the left and the right graph of each rule.
}
\label{fig:ruleComp:formoseExComp}
\end{figure*}

An important issue for the application to chemical reactions is that the
graphs involved in the rules are in general not connected.  Typical
chemical reactions combine molecules, split molecules or transfer groups of
atoms from one molecule to another.  The transformation rules for all these
reactions therefore require multiple connected components.  For the purpose
of dealing with these rules, we introduce the following notation for graphs
and derivations.

Let $Q$ be a graph with $\#Q$ connected components $Q_i$, $i=1,\dots,\#Q$.
It will be convenient to treat $Q$ as the multiset of its components. A
typical chemical graph derivation, corresponding to a bi-molecular reaction
can be written in the form $\{G^1,G^2\}\xRightarrow{p, m} \{H^1,H^2,H^3\}$,
where we take the notation to imply that all graphs $G^i$ and $H^j$ are
connected. We will furthermore insist that representations of chemical
reactions are minimal in the following sense: If the left graph of the rule
$p=(L\leftarrow K\rightarrow R)$ matches entirely within $G^1$, i.e.,
$m(L)\cap G^2=\emptyset$, then $G^2$ can be omitted. (In a chemical
rewriting grammar, then, one of the $H^i$ must be isomorphic to $G^2$,
becoming redundant as well.) More formally, we say that a derivation
$\{G^1, G^2,\dots, G^{\#G}\}\xRightarrow{p,m}\{H^1, H^2,\dots, H^{\#H}\}$ 
is \emph{proper} if
\begin{align*}
  \forall i, j : G_i\cong H_j\Rightarrow G_i\cap m(L)\ne\emptyset
\end{align*}
That is, a proper derivation cannot be simplified. If the derivation
$G\xRightarrow{p} H$ is proper then $\#G\leq\#L$.  The inequality comes
from that fact that multiple components of $L$ may easily be matched to a
single component of $G$ while each component of $L$ must match within a
component of $G$.

The conditions for the $\circ$ composition of rules are a bit too strict
for our applications. We thus relax them respect the component structure of
left and right graphs. More precisely, we require that $E$ is isomorphic to
a disjoint union of a copy of $R_1$ and some connected components of $L_2$
so that for every connected component $L^i_2$ of $L_2$ holds that either
$e_2(L^i_2)\subseteq e_1(R_1)$ or $e_2(L^i_2)$ is a connected component of
$E$ isomorphic to $L^i_2$. For a rule composition of this type to be well
defined we need that $\exists i$ such that $e_2(L_2^i)\subseteq e_1(R_1)$
holds. We remark that the latter condition could be relaxed further to lead to
additional compositions for which left and right sides are disjoint unions.

The composition of $p_1=(L_1,K_1,R_1)$ and $p_2=(L_2,K_2,R_2)$ now
yields $p_2\circ p_1 = (\{L_1,L^2_2\}, K_3, R_3)$ (cmp. Figure
\ref{fig:compnotsimp}). Note that right graph $R_3$ cannot no longer
be regarded simply as a rewritten version of $R_1$ because rule $p_2$
now adds additional vertices to both the left and the right graph. The
composite context $K_3$ contains only subsets of $K_1$ and $K_2$, but
it is expanded by the vertices of $L^2_2$ and the edges of $L^2_2$
that remain unchanged under rule $p_2$.

An example of a full rule composition is shown in
Fig.~\ref{fig:ruleComp:formoseExComp}.  The two rules in the example,
which in this case are also chemical reactions, are part of the Formose
grammar. The Formose grammar consists of two pairs of rules. The first pair
of rules, (from now on denoted as $\rr{p_0}$ and $\rr{p_1}$), implements
both directions of the keto-enol tautomerism. One direction, $\rr{p_1}$, is
visualized in Fig.~\ref{fig:ruleComp:formoseExComp}.  The second pair,
(from now on denoted as $\rr{p_2}$, $\rr{p_3}$) is the aldol-addition and
its reverse respectively. The reverse ($\rr{p_3}$) is also visualized in
Fig.~\ref{fig:ruleComp:formoseExComp}. We see that the left side of
$\rr{p_1}$ is isomorphic to a subgraph of one of the components of the
right side of $\rr{p_3}$.  Composing the two rules by subgraph matching
yields a third rule, $\rr{p_1} \circ \rr{p_3}$.

\begin{figure}
  \begin{center}
      \includegraphics[width=0.35\textwidth]{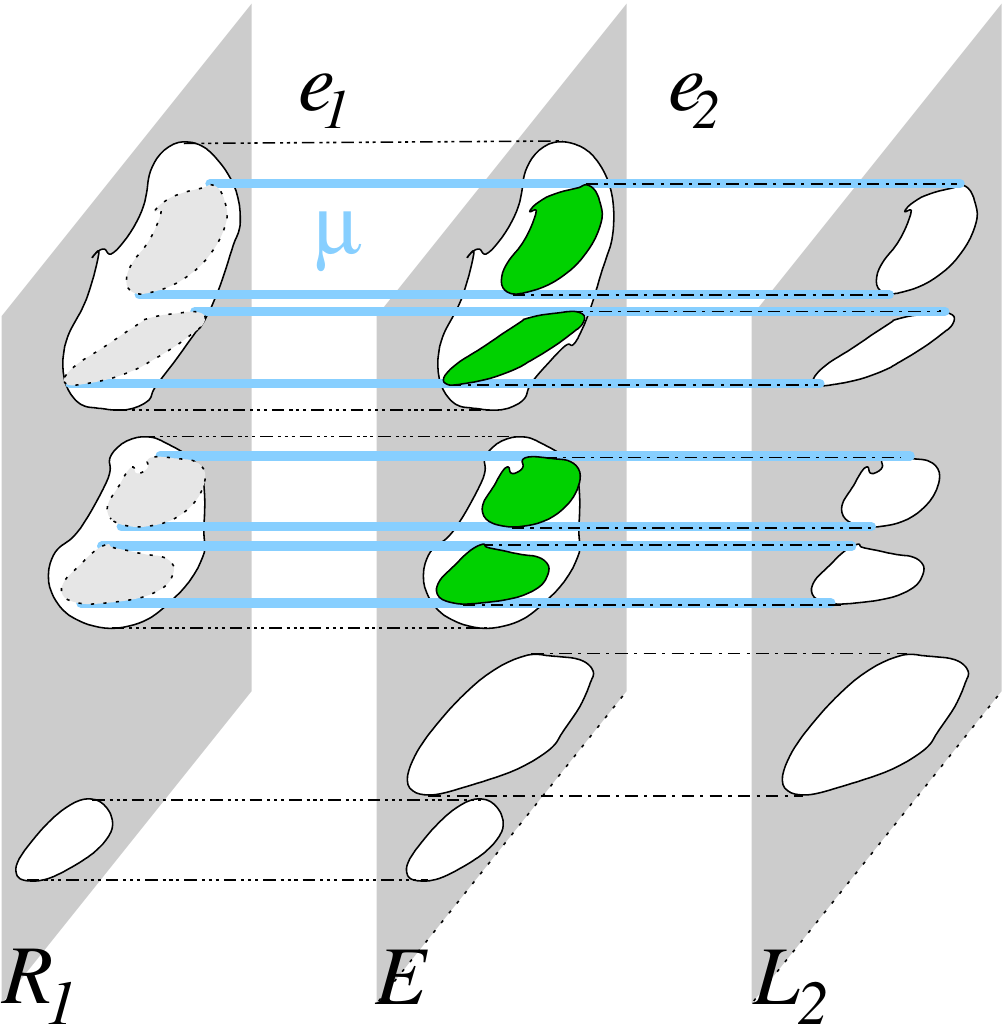}
      \caption{The (partial) composition of two rules is mediated by 
        the dependency graph $E$ and the two matching morphisms $e_1$ and
        $e_2$. Since these are subgraph isomorphisms in our case, $E$ is 
        simply the union $e_1(R_1)\cup e_2(L_2)$. The (partial) match
        $e_1(R_1)\cap e_2(L_2)$ can be understood as a matching $\mu$
        between $R_1$ and $L_2$, i.e., as a 1-1 relation of the matching 
        nodes \emph{and} edges. Whenever an edge is matched, then so are its
        incident vertices.}
      \label{fig:muh}
      \end{center}
\end{figure}

In general, we require here that the connected components of $R_1$ and
$L_2$ satisfy either $e_2(L^i_2)\subseteq e_1(R^j_1)$ or $e_1(R^j_1)\cap
e_2(L^i_2) =\emptyset$. We furthermore exclude the trivial case of parallel
rules in which only the second alternative is realized. In other extreme,
if all components $L_2^i$ satisfy $e_1(L^i_2)\subseteq
e_2(R_1)$, the partial composition becomes a full composition. Formally,
these alternatives are described by different dependency graphs $E$ and/or
different morphisms $e_1$ and $e_2$. Pragmatically we can understand this
as a matching $\mu$ of $L_2$ and $R_1$ as in Fig.~\ref{fig:muh}. Specifying
$\mu$ of course removes the ambiguity from the definition of the rule
composition; hence we write $p_2\circ_{\mu} p_1$ to emphasize the matching
$\mu$.

\section{Constructing Rule Compositions}
\label{sec:cons}
Given two rules, $p_1$ and $p_2$, it is not only interesting to know if a
partial composition is defined, but also to create the set of all possible
compositions
\begin{align*}
  \left\{p_2\circ_{\mu_1} p_1, p_2\circ_{\mu_2} p_1, \dots,
    p_2\circ_{\mu_k}p_1\right\}
\end{align*}
explicitly. This set in particular contains also all full compositions.
The following describes an algorithm for enumerating all
partial compositions.

\subsection{Enumerating the Matchings $\mu$}
The key to finding all compositions is the enumeration of all matchings
$\mu$ that respect out restrictions on overlaps between connected
components. We thus start from the sets $\left\{R_1^1, R_1^2, \dots,
  R_1^{\#R_1}\right\}$ and $\left\{L_2^1, L_2^2, \dots,
  L_2^{\#L_2}\right\}$ of connected components of $R_1$ and $L_2$, resp.
In the first set we find all subgraph matches $L_2^i\subseteq R_1^j$
(represented as the corresponding matchings $\mu_{ij}$) and arrange the
result in a matrix of lists of subgraph matches,
Fig.~\ref{fig:ruleComp:matchMatrix}.

\begin{figure}[b]
\begin{tabular}{lcr}
\begin{minipage}{0.4\columnwidth}
\centering \subfloat[Match matrix]{
\begin{tabular}{c|*{3}{p{0.5cm}|}}
  \multicolumn{1}{c}{}
  & \multicolumn{1}{c}{$R_1^1$}
  & \multicolumn{1}{c}{$R_1^2$}
  & \multicolumn{1}{c}{$R_1^3$}	\\ \cline{2-4}
  $L_2^1$	& 1	&	& 2	\\ \cline{2-4}
  $L_2^2$	&	& 1	& 1	\\ \cline{2-4}
\end{tabular}
\label{fig:ruleComp:matchMatrix}
}
\end{minipage}
&& 
\begin{minipage}{0.5\columnwidth}
\subfloat[Extended match matrix]{
\begin{tabular}{c|*{4}{p{0.5cm}|}}
  \multicolumn{1}{c}{}
  & \multicolumn{1}{c}{$R_1^1$}
  & \multicolumn{1}{c}{$R_1^2$}
  & \multicolumn{1}{c}{$R_1^3$}	
  & \multicolumn{1}{c}{$R_1^\emptyset$}		\\ \cline{2-5}
  $L_2^1$	& 1	&	& 2	& 1	\\ \cline{2-5}
  $L_2^2$	&	& 1	& 1	& 1	\\ \cline{2-5}
\end{tabular}
\label{fig:ruleComp:matchMatrixExt}
}
\end{minipage}
\end{tabular}
\caption{Example of a match matrix and the same matrix with its virtual
  extension.  The top row specifies 1 possibility for $L_2^1\subseteq
  R_1^1$ and 2 for $L_2^1\subseteq R_1^3$.  The extended matrix further
  specifies that $L_2^1$ can be unmatched.  The bottom rows can be
  interpreted similarly. We display the number of matchings instead of a
  representation of the matchings themselves. }
\label{fig:ruleComp:matchMatrixBoth}
\end{figure}

The matching matrix is extended by a virtual column to account for the
possibility that $L_2^i$ is not matched with any component of $R_1$.  Every
partial (and full) composition is now defined by a selection of one
submatch from each row of the matrix, see Supplemental Material for an
example. The converse is not true, however: Not every selection of matches
correspond to a partial composition. In particular, we exclude the case
that only entries from the virtual column are selected. In addition, the
sub-matches must be disjoint to ensure that the combined match is
injective. The latter conditions needs to be checked only when more than
one submatch is selected from the same column.

\subsection{Composing the Rules}

The construction of the composition $p_2\circ_\mu p_1$ of two rules $p_1$
and $p_2$ does not explicitly depend on the component structure of $R_2$
and $L_1$ because it is uniquely defined by the matching $\mu$ and the
bijections of the nodes of $L_i$, $K_i$, and $R_i$ for each of the two
rules. We obtain $L$ by extending $L_1$ with unmatched components of $L_2$
and $R$ by extending $R_2$ by the unmatched components of $R_1$.  The
corresponding extension of $\mu$ to a bijection $\hat\mu$ of the vertex
sets of $L$ and $R$ is uniquely defined. The context $K$ of the composite
rule simply consists the common vertex set of $L$ and $R$ and all edges
$(x,y)$ of $L$ for which $(\hat\mu(x),\hat\mu(y))$ is an edge in $R$. We
note in passing that $\hat\mu$ defines the atom mapping of the composite
transformation. The explicit construction of $(R,K,L)$ is summarized as
Algorithm~\ref{alg:ruleComp}.

\begin{algorithm}[t]
\dontprintsemicolon
\KwIn{$p_1 = (L_1, K_1, R_1)$}
\KwIn{$p_2 = (L_2, K_2, R_2)$}
\KwIn{$\mu$, a partial matching between $L_2$ and $R_1$}
\KwOut{$p  = (L, K, R)$}
$p\leftarrow$ empty rule\;
Copy vertices of $p_1$ to $p$\;
\ForEach{vertex $v\in p_2$}{
	\If{$v$ is not mapped by $\mu$} {
		Copy $v$ to $p$
	}\Else{
		Change membership in $L$, $K$ and $R$ for vertex $\mu(v)$
	}
}
Copy edges of $p_1$ to $p$\;
\ForEach{Edge $e\in p_2$}{
	\If{$e$ is not mapped by $\mu$} {
		Copy $e$ to $p$
	}\Else{
		Change membership in $L$, $K$ and $R$ for edge $\mu(e)$
	}
}
Delete edges and vertices created by $p_1$, but deleted by $p_2$\;
\lIf{matching condition not satisfied}{\Abort}\;
\Return{$p$}
\caption{Composing $p_1$ and $p_2$ to $p$, by a given partial mapping}
\label{alg:ruleComp}
\end{algorithm}
The implementation of the algorithm naturally depends heavily on the
representation of transformation rules, which in our implementation is the
representation from the Graph Grammar Library (GGL) \cite{xtofEvo}. The
representation is a single graph, with attached vertex and edge properties
defining membership of $L$, $K$ and $R$, as well as the needed labels.

\begin{figure*}[t]
\newcommand{\ruleFig}[3]{%
\includegraphics[width=#3\textwidth]{#1_#2}
}
\newcommand{\ruleFigLeft}[1]{\ruleFig{#1}{left}{0.28}}
\newcommand{\ruleFigRight}[1]{\ruleFig{#1}{right}{0.56}}
\begin{center}
\begin{tabular}{cccc}
\begin{minipage}[t]{0.375\textwidth}
\includegraphics[width=\textwidth]{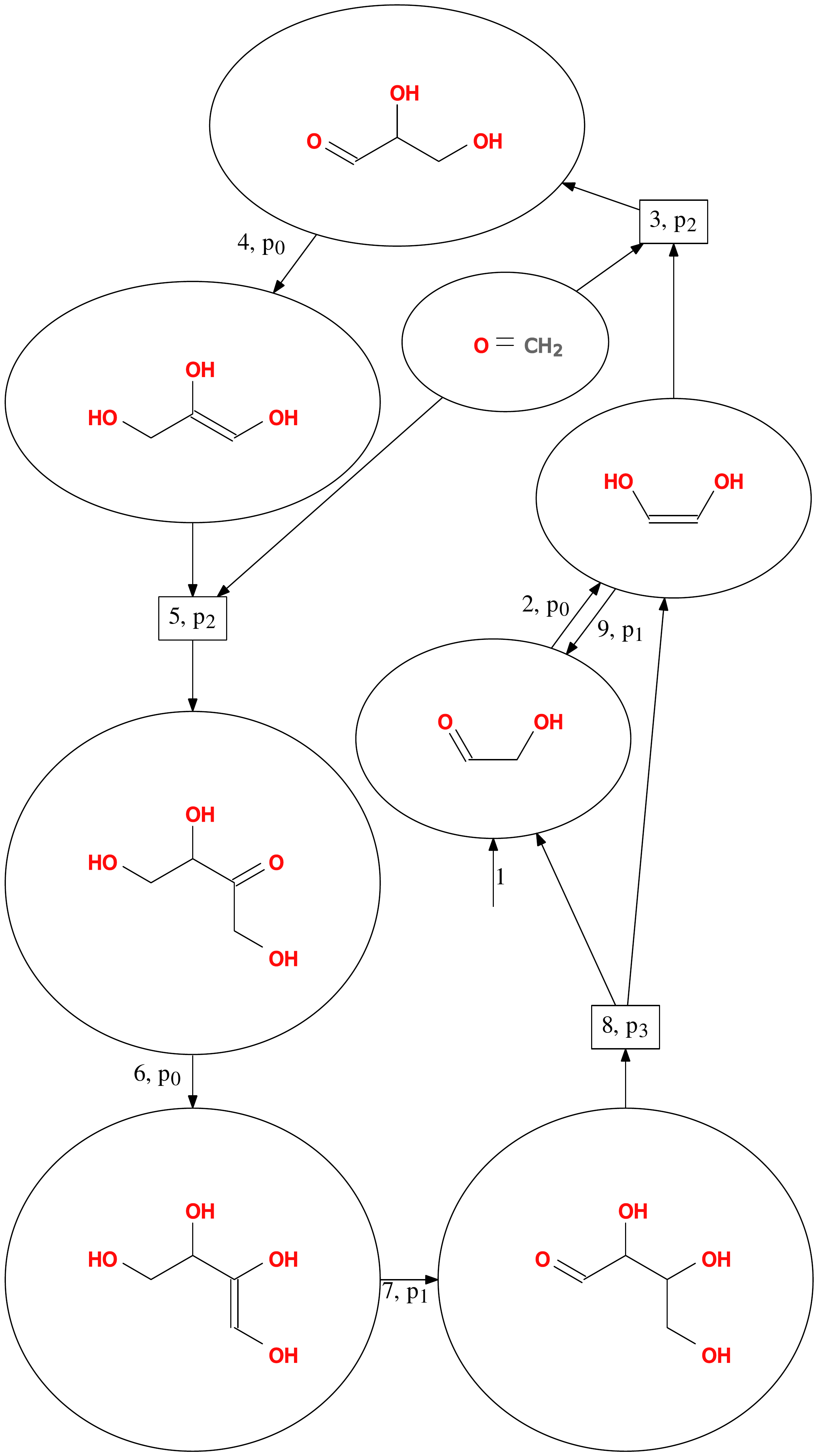}\\
\end{minipage}
& \qquad & 
\begin{minipage}[t]{0.25\textwidth}
\begin{tikzpicture}[
node distance=0cm
]
\node[draw] (Right1) {\ruleFigRight{1}};
\node[below,draw] (Right2) at (Right1.south) {\ruleFigRight{2}};
\node[below,draw] (Right3) at (Right2.south) {\ruleFigRight{3}};
\node[below,draw] (Right4) at (Right3.south) {\ruleFigRight{4}};
\node[below,draw] (Right5) at (Right4.south) {\ruleFigRight{5}};
\foreach \i in {1, ..., 5}{
	\node[draw] (Left\i) [left=of Right\i ] {\ruleFigLeft{\i}};
	\node[] (L\i) [left=of Left\i] {$\i$};
}
\node[below] at (Left5.south) {$L_i$};
\node[below] at (Right5.south) {$R_i$};
\end{tikzpicture}
\end{minipage}
&
\begin{minipage}[t]{0.25\textwidth}
\begin{tikzpicture}[
node distance=0cm
]
\node[draw] (Right6) {\ruleFigRight{6}};
\node[below,draw] (Right7) at (Right6.south) {\ruleFigRight{7}};
\node[below,draw] (Right8) at (Right7.south) {\ruleFigRight{8}};
\node[below,draw] (Right9) at (Right8.south) {\ruleFigRight{9}};
\foreach \i in {6, ..., 9}{
	\node[draw] (Left\i) [left=of Right\i ] {\ruleFigLeft{\i}};
	\node[] (L\i) [left=of Left\i] {$\i$};
}
\node[below] at (Left9.south) {$L_i$};
\node[below] at (Right9.south) {$R_i$};

\end{tikzpicture}
\end{minipage}
\end{tabular}
\end{center}
\caption{Above: Chemical reaction network for the Formose reaction;
  hyperedges are labeled with $(i,\rr{p_j})$ where $i$ is the $i$-th
  reaction $\rho_i$ in the rule composition.  $\rr{p_j}, 0\leq j \leq 3$,
  refers to a specific rule from the Formose reaction; Right: Resulting
  composed rule after the composition of the first $i$ rules along the
  Formose cycle, context shown in black.}
\label{fig:res}
\end{figure*}

Not all matchings define valid rule composition. For instance, consider an
edge $(u, v)$ that is present in $R_1$ and $R_2$ but not in $L_2$ and both
$u$ and $v$ are in $L_2$. This would amount to creating the edge by means
of rule $p_2$ which was already introduced by $p_1$. Since we do not allow
parallel edges and thus regard such inconsistencies as undefined cases and
reject the matching.  Note that a parallel edge does not correspond to a
``double bond'' (which essentially is only an edge with a specific type).

\subsection{Graph Binding}
The composition of transformation rules, and thereby chemical reactions,
makes it possible to create abstract meta-rules in a way that is similar to
the combination of multiple functions into more abstract functions in
functional programming. A related concept from (functional)
programming that seems useful in the context of graph grammars is partial
function application.  Consider, for example, the binding of the number $2$
to the exponentiation operator, yielding either the function $f(x) = 2^x$
or $f(x) = x^2$.  In the framework of rule composition, we define graph
binding as a special case.

Let $G$ be a graph and $p_2 = (L_2, K_2, R_2)$ be a transformation rule.
The binding of $G$ to $p_2$ results in the transformation rule $p = (L, K,
R)$ which implements the partial application of $p_2$ on $G$.  This is
accomplished simply by regarding $G$ as a rule $p_1 = (\emptyset,
\emptyset, G)$, and using partial composition; $p = p_2\circ p_1$.
Note that if $p_2\circ p_1$ is a full composition, then $p$ can be regarded
as a graph $H$ and $G\xRightarrow{p_2}H$ holds.

Graph binding allows a simplified representation of reactions.  For
instance, we can use this formal construction to omit uninteresting
ubiquitously present molecules such as water by binding the graph of the water
molecule to the transformation rule of a reaction that requires water.
Similarly, graph unbinding can be defined as a transformation rule that
destroys graphs. In a chemical application it can be used to avoid the
explicit representation of uninteresting ubiquitous molecules such as the
solvent.

\subsection{Ordering Rules}
\label{subsec:ordering}
A wide variety of methods, including flux balance analysis, can be used to
identify pathways or other subsets of reactions that are of
interest. Adjacency of reactions in the original networks as well as their
directionality can be used efficiently to prune the possible orders of rule
compositions. The fact that multiple reactions are instantiations of the
same transformation rule, as in the example discussed in detail in the next
section, further reduces the search spaces.

\section{Results and Discussion}
\label{sec:res}

We illustrate the use of transformation rule composition by deriving of
meta-rules from the graph grammar consisting of the four rules necessary to
represent the complete Formose reaction, see
Fig.~\ref{fig:ruleComp:formoseExComp}. The overall reaction pattern of the
Formose cycle is $2 g_0 + g_1 \rightarrow 2 g_1$ with $g_0$ being
formaldehyde and $g_1$ being glycolaldehyde. It amounts to the linear
combination $\sum_{i=1}^{9} \rho_i$ of the eight reactions and the influx
$\rho_1$ of $g_0$ listed in Fig.~\ref{fig:ruleComp:formoseExComp}. It is
important to notice that several of these reactions are instantiations of
the same, well-known chemical transformations. We have forward keto-enol
tautomerism ($\rr{p_0}$: $\rho_2$, $\rho_4$, $\rho_6$), backward keto-enol
tautomerism ($\rr{p_1}$: $\rho_7$, $\rho_9$), forward aldol addition
($\rr{p_2}$: $\rho_3$, $\rho_5$), and backward aldol addition ($\rr{p_3}$:
$\rho_8$). The composite rule models the complete autocatalytic cycle shown
in Fig.~\ref{fig:res} as a single meta-rule.

Throughout this section we will not explicitly distinguish between partial
composition and full composition, and we interpret the composition operator
$\circ$ as right-associative to simplify the notation. Thus $p_i\circ
p_j\circ p_k$ means $p_i \circ (p_j\circ p_k)$.

The rules are used in the autocatalytic cycle in the following order
(starting with an keto-enol tautomerisation $\rr{p_0}$):
\begin{align*}
\rr{p_0, p_2, p_0, p_2, p_0, p_1, p_3, p_1}
\end{align*}
As it is not possible to compose this sequence of rules directly, we start
by binding glycolaldehyde $g_1$ to reaction $p_0$, as the before-mentioned
keto-enol tautomerisation is applied to molecule $g_1$. The resulting rule
is denoted as $\rr{g_1}$. The hyperedges in the chemical reaction network
depicted in Fig.~\ref{fig:res} are numbered according to the sequence that
reflects in which order the Formose reaction takes place and consequently
the order in which the rule composition subsequently is done. The first
composition refers to the binding operation. This binding of glycolaldehyde
results in a graph grammar rule, which is depicted in row 1 in the table
depicted in Fig.~\ref{fig:res}, i.e., the rule ($\emptyset$, $\emptyset$,
$g_1$) (see ``Graph Binding''). The numbers at the hyperedges ($2,3, \ldots
9$) refer to the second, third, $\ldots$, ninth reaction in the sequence of
reactions given above. The graph grammar rule $\rr{p_i}, 0 \leq i \leq 3$,
used for the corresponding hyper-edge is given next to the sequence
number. The rules inferred by a subsequent rule composition are given in
rows $2$ to $9$ of the table.

The application of the final rule results in the composed meta-rule
$\rr{p_1} \circ \rr{p_3} \circ \ldots \circ \rr{p_0} \circ \rr{g_1}$. This
rule precisely covers the reaction pattern of the Formose reaction, namely
how two formaldehyde molecules and one (bound) glycolaldehyde are
transformed to two glycolaldehyde molecules. However note, that the rule is
general enough such that any pair of molecules with aldehyde groups can be
used, i.e., the inferred reaction pattern refers to a class of overall
reactions and the product does not necessarily need to be glycolaldehyde.

The practical computation of these compositions takes less than a second in
the current implementation.  Even for substantially more general
composition sequences the running time remains manageable. For instance, it
takes less than $1$ minute to compute all composition sequences with a
length $k\le 10$ of the form $p_{i_1} \circ p_{i_2} \circ \dots \circ
p_{i_k} \circ g_q$ with $i_j\in \{0, 1, 2, 3\}$, based on the binding of
one of the influx molecules $g_0$ or $g_1$. This results in 1875 different
inferred composite rules.

Polymerization can also be viewed as a pathway in a chemical reaction
network, albeit one of potentially infinite size. The same methods applied
to the automatic inference of the overall reaction pattern of the Formose
cycle can be directly applied to detecting composition rules for
polymerization reactions. Importantly, even if a chemical reaction network
is not given, the approaches presented in this paper can be used to
automatically find sequences of reactions that will lead to
polymerization. This can be realized by a straight-forward post-processing
step: all that needs to be done is to check whether an inferred composite
rule exhibits a replicated functional unit. Such polymerization meta-rules
also enable the analysis of chemical systems with highly complex carbon
skeletons such as the natural compound classes of the terpenes or the
polyketides.

\section{Conclusions}

Graph grammars provide a convenient framework for modeling chemistries on
different levels of abstraction. A chemically valid approach is to see any
chemical reaction as a bi-molecular reaction. This requires graph grammar
rules that cover changes of molecules in an rather explicit and detailed
way. Understanding chemical reaction patterns usually requires spanning the
chemical reaction networks based on such rules. Obviously, this approach
suffers the inherent potential of an immense combinatorial explosion. In
this paper we introduced the automatic inference of such higher-level
chemical reaction pattern based on a formal approach for graph grammar rule
combination. We analyzed the autocatalytic cycle of the Formose reaction
and inferred its overall reaction pattern as a rule composition of nine
rules. Rule composition is also naturally applicable to inferring patterns
of polymerization reactions. Future work will include e.g.\ the analysis of
terpene-based and hydrogen cyanide-based polymerization chemistry.

\subsection*{Acknowledgments} 
This work was supported in part by the Volkswagen Stiftung
proj. no. I/82719, and the COST-Action CM0703 ``Systems Chemistry'' and by
the Danish Council for Independent Research, Natural Sciences.

\label{sec:con}
\bibliographystyle{plainnat} 
\bibliography{BioInfo-2012-RuleComp-TR}
\newcounter{endmaintextpage}
\setcounter{endmaintextpage}{\value{page}}
\clearpage
\pagenumbering{roman}
\setcounter{page}{1}
\newpage
\appendix
\section*{Supplemental Material:\newline
Example of Enumeration of Compositions}
In this Appendix we show the complete result of the composition of two
(artificial) rules, $p_1$ and $p_2$, including the selection of submatches
from the match matrix. The two rules are depicted in
Fig.~\ref{fig:appExample} with the extended match matrix of the composition
$p_2\circ p_1$, that corresponds to the example of an extended match matrix
as given in the paper. The rules in this section are all depicted with
vertices that have an additional index. The numbering of the components is
in increasing order wrt.\ to these indices, e.g., $L_2^1$ denotes the
component connecting nodes $A,0$ and $B,1$ and $L_2^2$ denotes the
component connecting nodes $B,2$ and $C,3$.  
\newcommand\compScale{0.2}
\newcommand{\insertComp}[1]{
\begin{figure}[h]
\begin{center}
\fbox{
\includegraphics[scale=\compScale]{res#1_left}
}
\fbox{
\includegraphics[scale=\compScale]{res#1_right}
}
\end{center}
\centerline{Result of composition #1}
\end{figure}
}

\begin{figure}[h]
\begin{center}
\subfloat[$p_1$]{
\fbox{
\includegraphics[scale=\compScale]{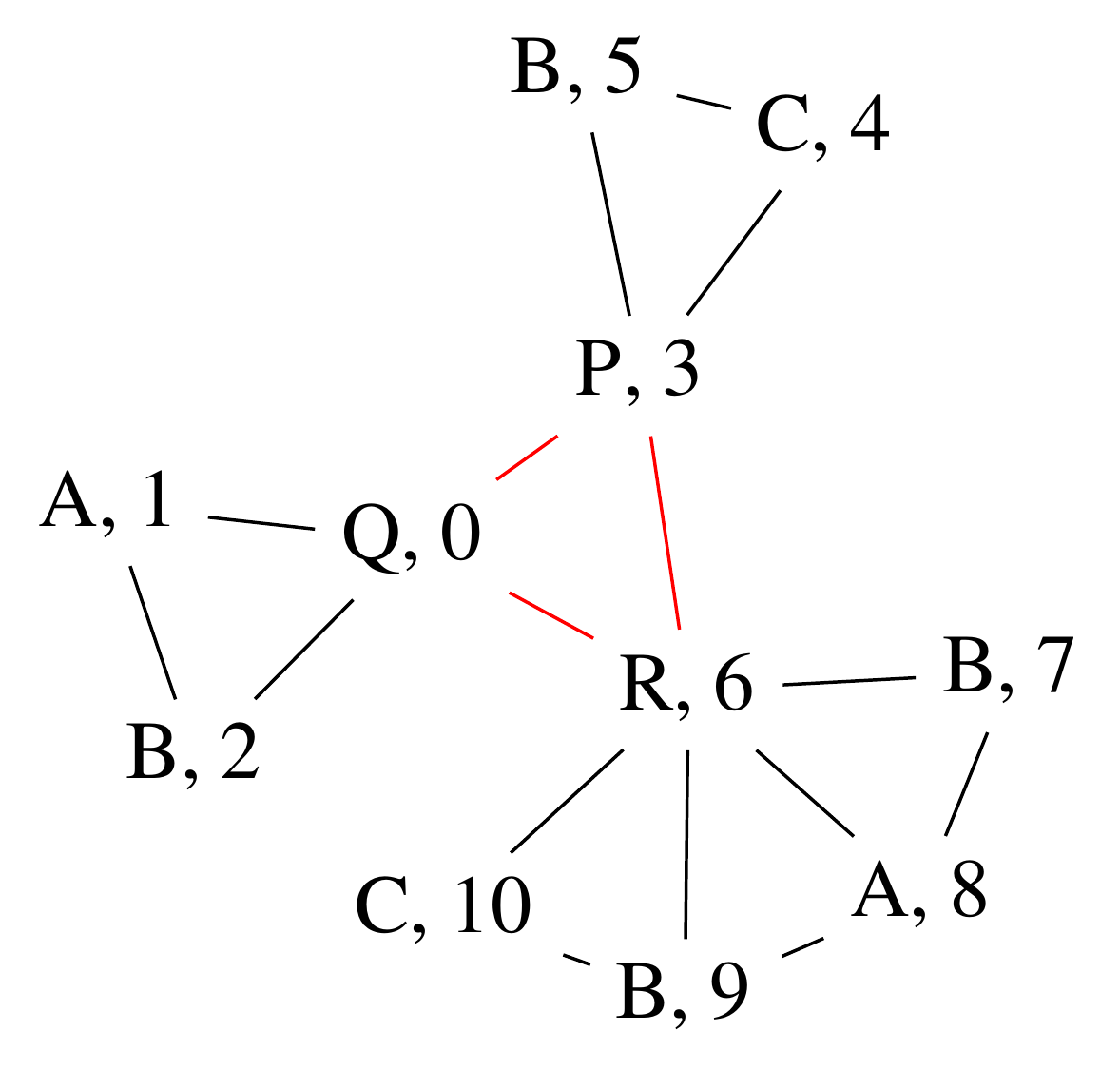}
}
\fbox{
\includegraphics[scale=\compScale]{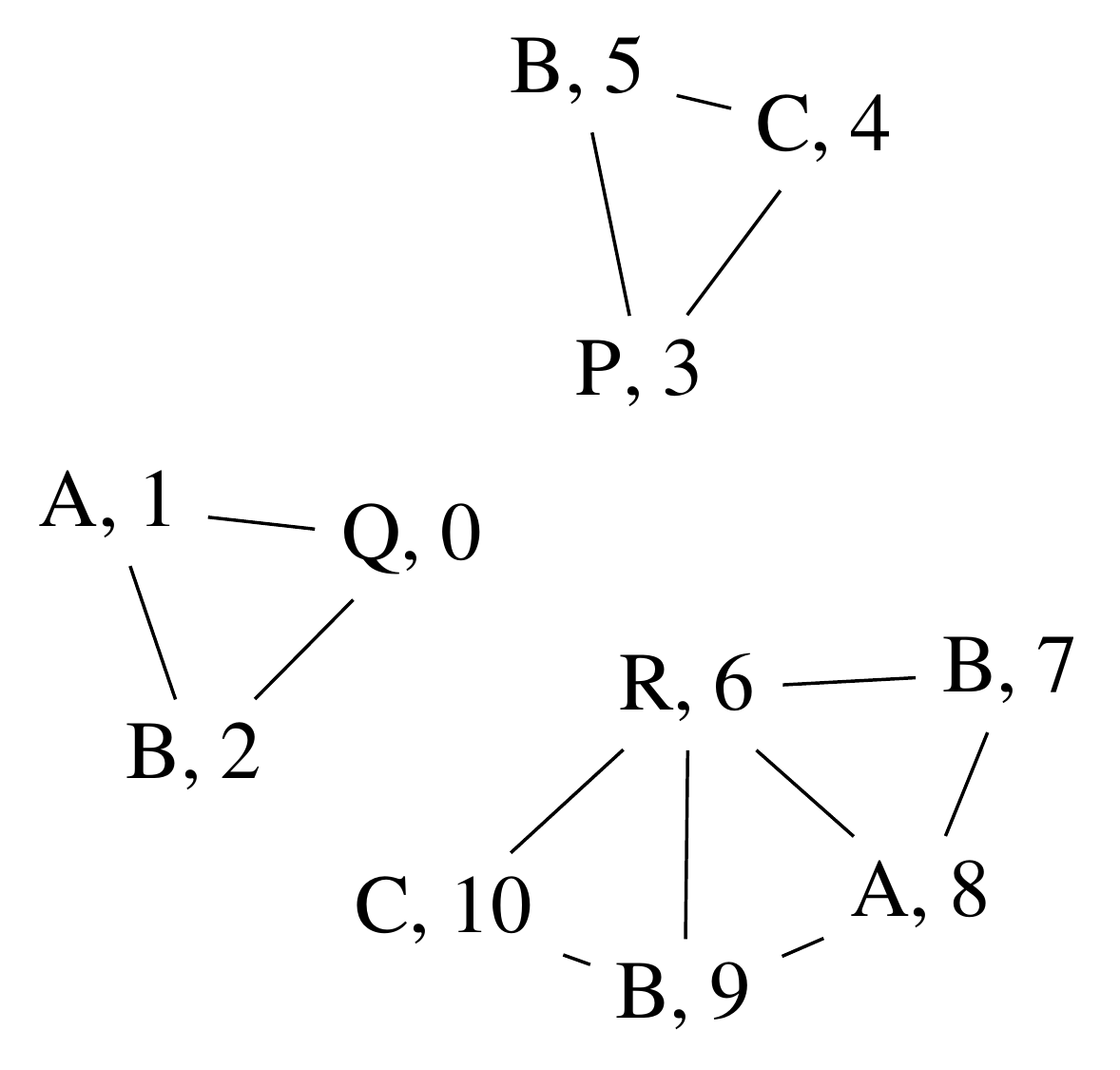}
}
}
\subfloat[$p_2$]{
\fbox{
\includegraphics[scale=\compScale]{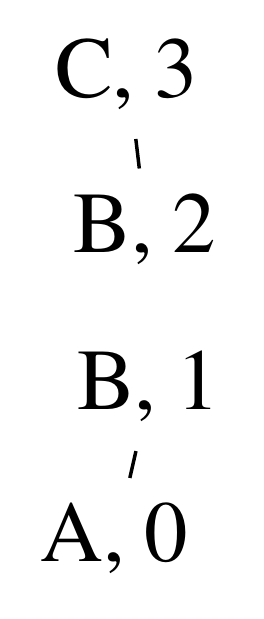}
}
\fbox{
\includegraphics[scale=\compScale]{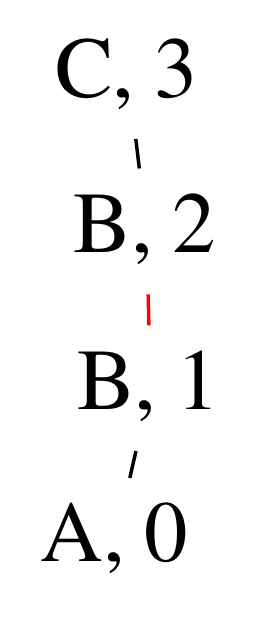}
}
}
\\
\subfloat[Extended match matrix]{
\hspace{1cm}
\begin{tabular}{c|*{4}{p{0.5cm}|}}
  \multicolumn{1}{c}{}
  & \multicolumn{1}{c}{$R_1^1$}
  & \multicolumn{1}{c}{$R_1^2$}
  & \multicolumn{1}{c}{$R_1^3$}	
  & \multicolumn{1}{c}{$R_1^\emptyset$}		\\ \cline{2-5}
  $L_2^1$	& 1	&	& 2	& 1	\\ \cline{2-5}
  $L_2^2$	&	& 1	& 1	& 1	\\ \cline{2-5}
\end{tabular}
\hspace{1cm}
}
\end{center}
\caption{The two rules $p_1$ and $p_2$, and the extended match matrix of the composition $p_2\circ p_1$.
The components of both $R_1$ and $L_2$ are numbered in the same order as the vertex indices. 
}
\label{fig:appExample}
\end{figure}

In the following we will enumerate all valid selections of submatches based on the extended match matrix and give the corresponding resulting rule composition. The chosen matches are depicted as \textbullet~in the extended match matrix. If several matches can be found (in our example this is true for the component $L_2^1$, that can be matched twice in $R_1^3$), the \textbullet~has an index.
\subsection*{Composition 1}
\begin{center}
\begin{tabular}{c|*{4}{c|}}
  \multicolumn{1}{c}{}
  & \multicolumn{1}{c}{$R_1^1$}
  & \multicolumn{1}{c}{$R_1^2$}
  & \multicolumn{1}{c}{$R_1^3$}	
  & \multicolumn{1}{c}{$R_1^\emptyset$}		\\ \cline{2-5}
  $L_2^1$	& \textbullet	&	& 	& 	\\ \cline{2-5}
  $L_2^2$	&	& \textbullet	& 	& 	\\ \cline{2-5}
\end{tabular}
\end{center}
\insertComp{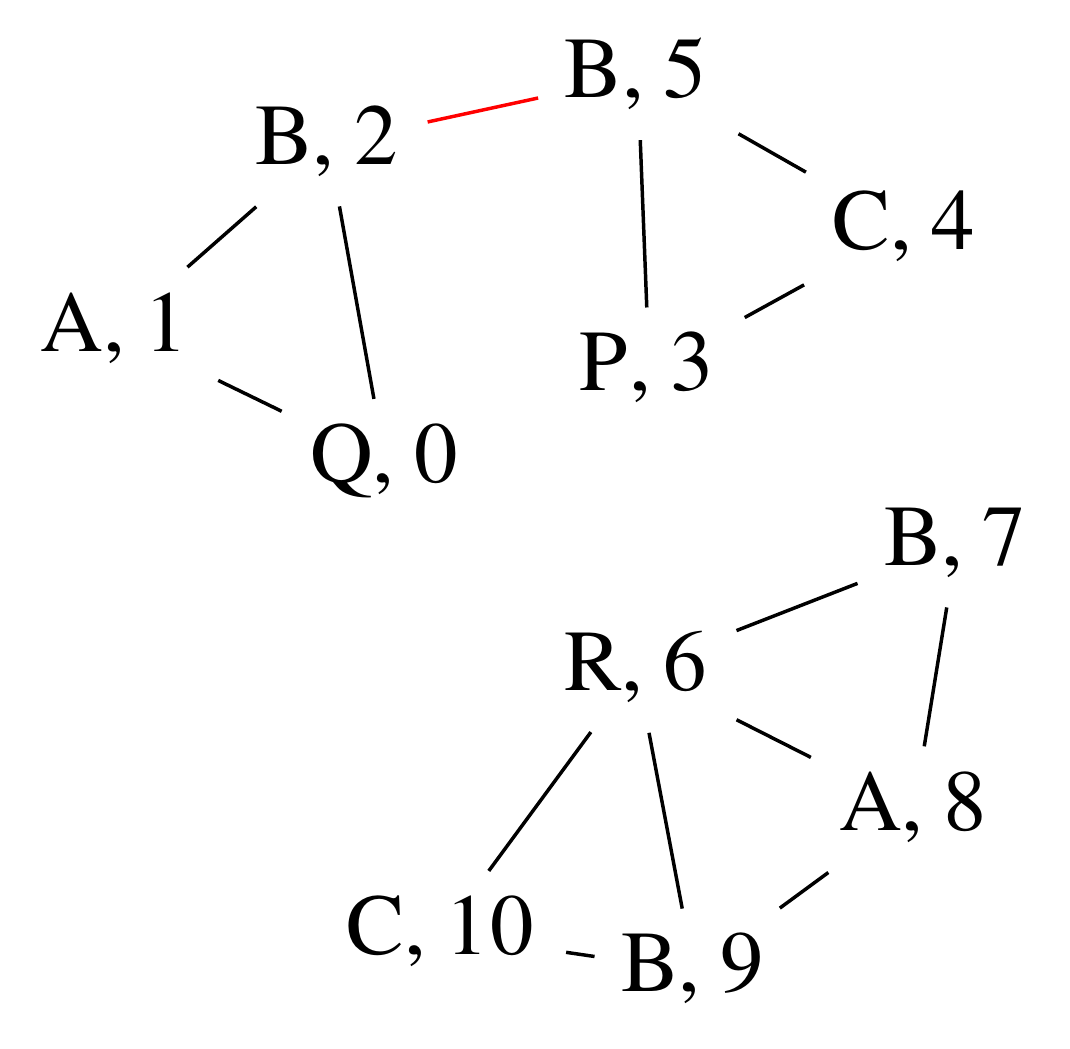}

\subsection*{Composition 2}
\begin{center}
\begin{tabular}{c|*{4}{c|}}
  \multicolumn{1}{c}{}
  & \multicolumn{1}{c}{$R_1^1$}
  & \multicolumn{1}{c}{$R_1^2$}
  & \multicolumn{1}{c}{$R_1^3$}	
  & \multicolumn{1}{c}{$R_1^\emptyset$}		\\ \cline{2-5}
  $L_2^1$	& 	&	& \textbullet$_1$	& 	\\ \cline{2-5}
  $L_2^2$	&	& \textbullet	& 	& 	\\ \cline{2-5}
\end{tabular}
\end{center}
\insertComp{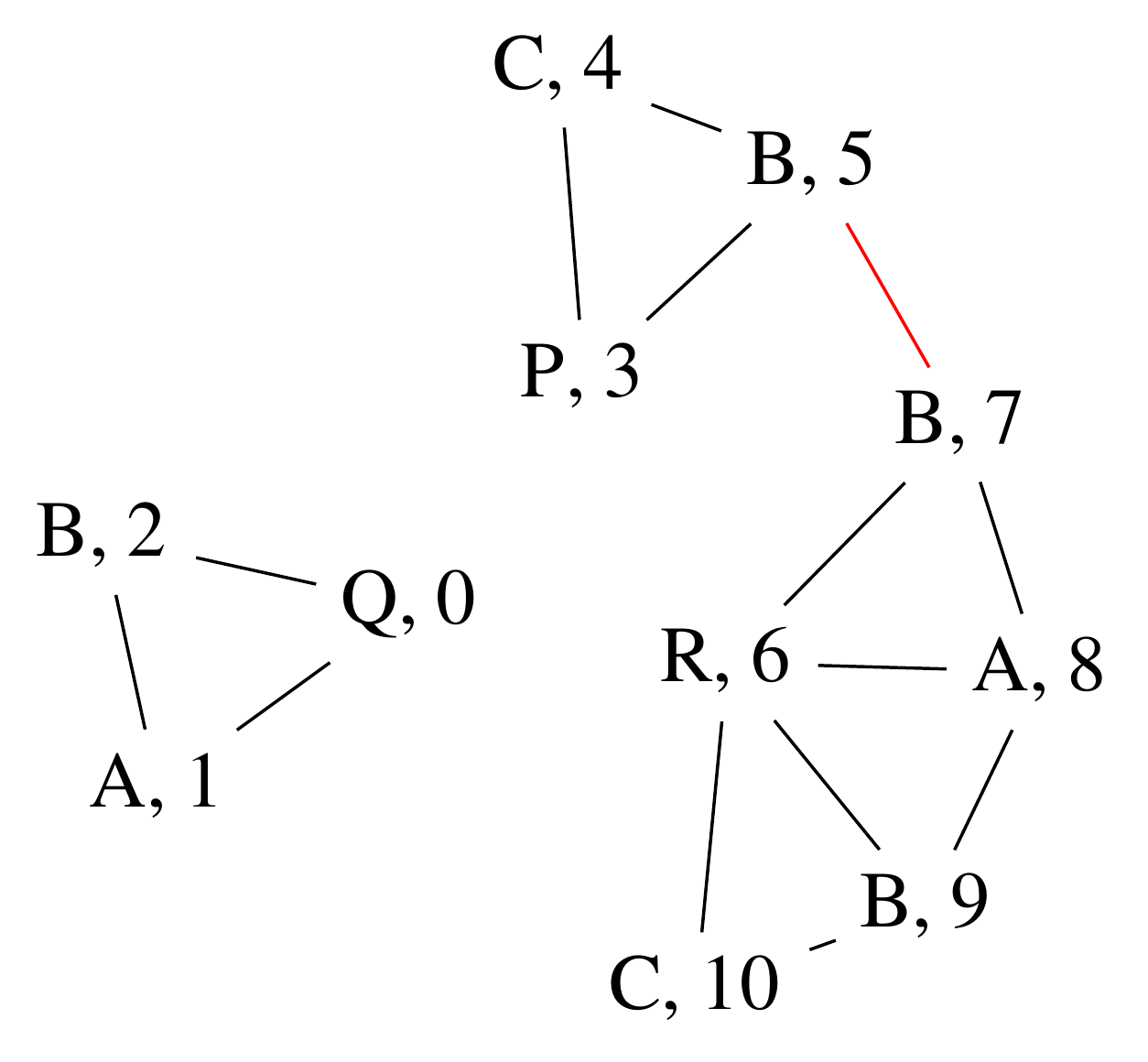}

\subsection*{Composition 3}
\begin{center}
\begin{tabular}{c|*{4}{c|}}
  \multicolumn{1}{c}{}
  & \multicolumn{1}{c}{$R_1^1$}
  & \multicolumn{1}{c}{$R_1^2$}
  & \multicolumn{1}{c}{$R_1^3$}	
  & \multicolumn{1}{c}{$R_1^\emptyset$}		\\ \cline{2-5}
  $L_2^1$	& 	&	& \textbullet$_2$	& 	\\ \cline{2-5}
  $L_2^2$	&	& \textbullet	& 	& 	\\ \cline{2-5}
\end{tabular}
\end{center}
\insertComp{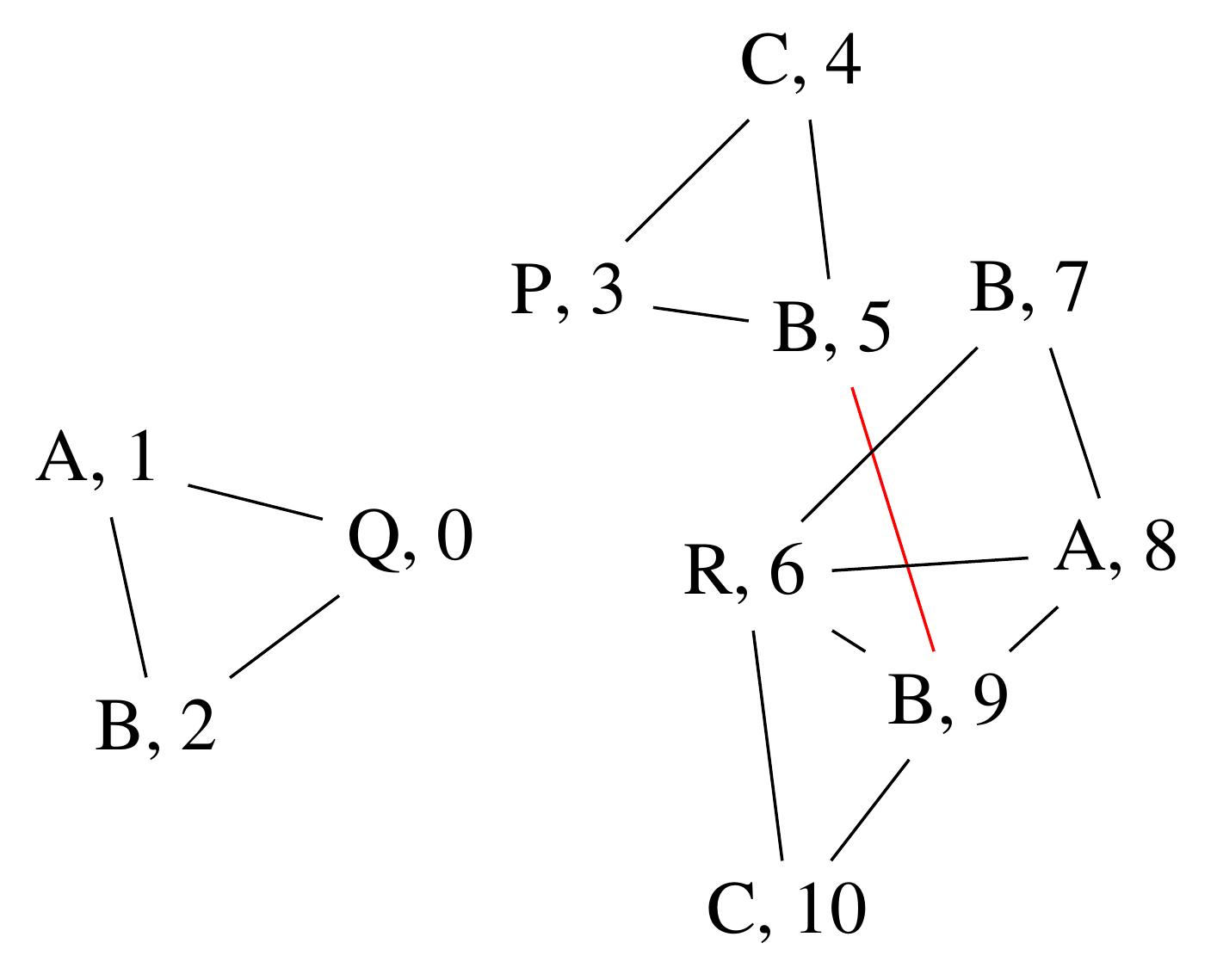}

\subsection*{Composition 4}
\begin{center}
\begin{tabular}{c|*{4}{c|}}
  \multicolumn{1}{c}{}
  & \multicolumn{1}{c}{$R_1^1$}
  & \multicolumn{1}{c}{$R_1^2$}
  & \multicolumn{1}{c}{$R_1^3$}	
  & \multicolumn{1}{c}{$R_1^\emptyset$}		\\ \cline{2-5}
  $L_2^1$	& 	&	& 	& \textbullet	\\ \cline{2-5}
  $L_2^2$	&	& \textbullet	& 	& 	\\ \cline{2-5}
\end{tabular}
\end{center}
\insertComp{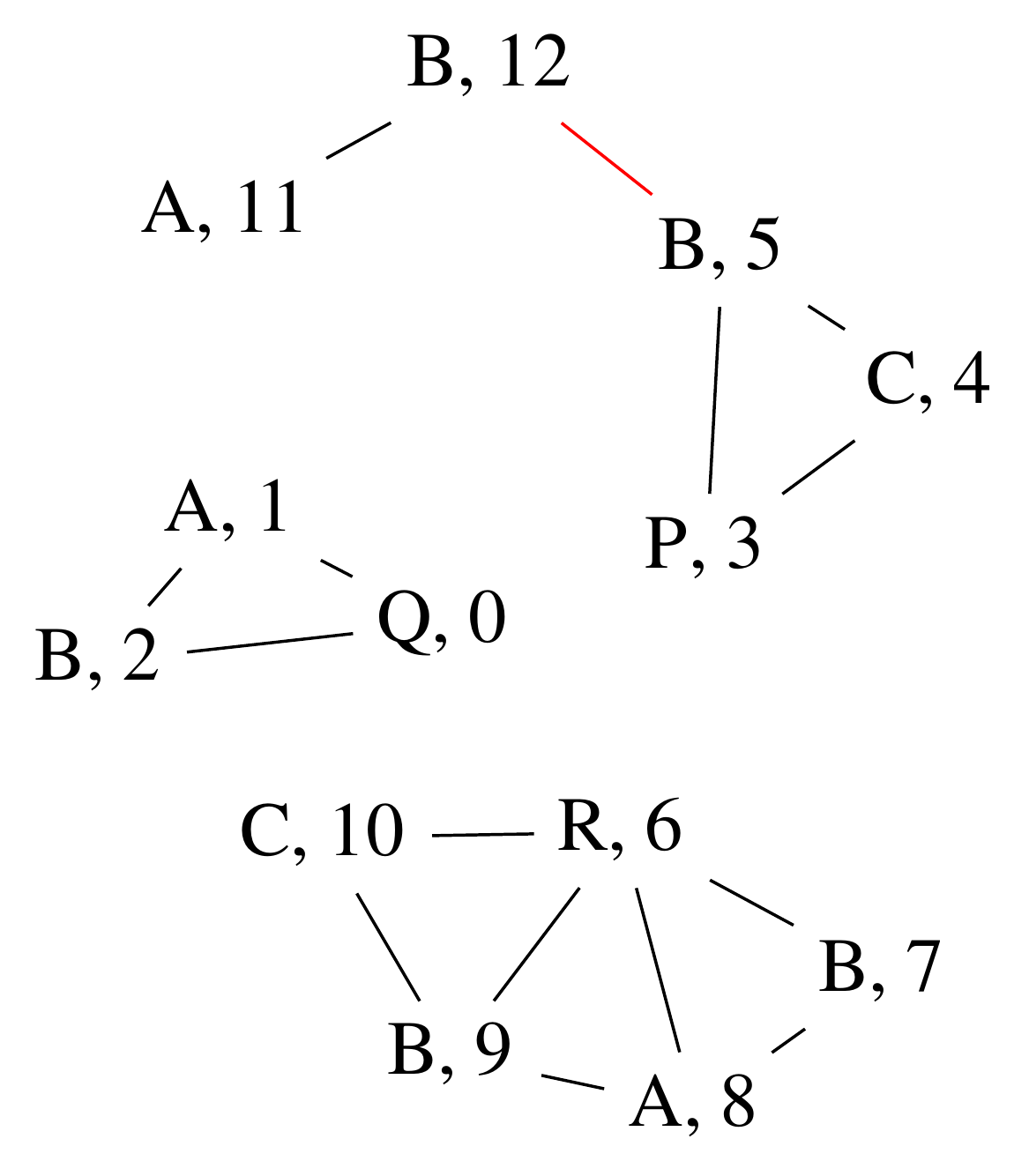}

\subsection*{Composition 5}
\begin{center}
\begin{tabular}{c|*{4}{c|}}
  \multicolumn{1}{c}{}
  & \multicolumn{1}{c}{$R_1^1$}
  & \multicolumn{1}{c}{$R_1^2$}
  & \multicolumn{1}{c}{$R_1^3$}	
  & \multicolumn{1}{c}{$R_1^\emptyset$}		\\ \cline{2-5}
  $L_2^1$	& \textbullet	&	& 	& 	\\ \cline{2-5}
  $L_2^2$	&	& 	& \textbullet	& 	\\ \cline{2-5}
\end{tabular}
\end{center}
\insertComp{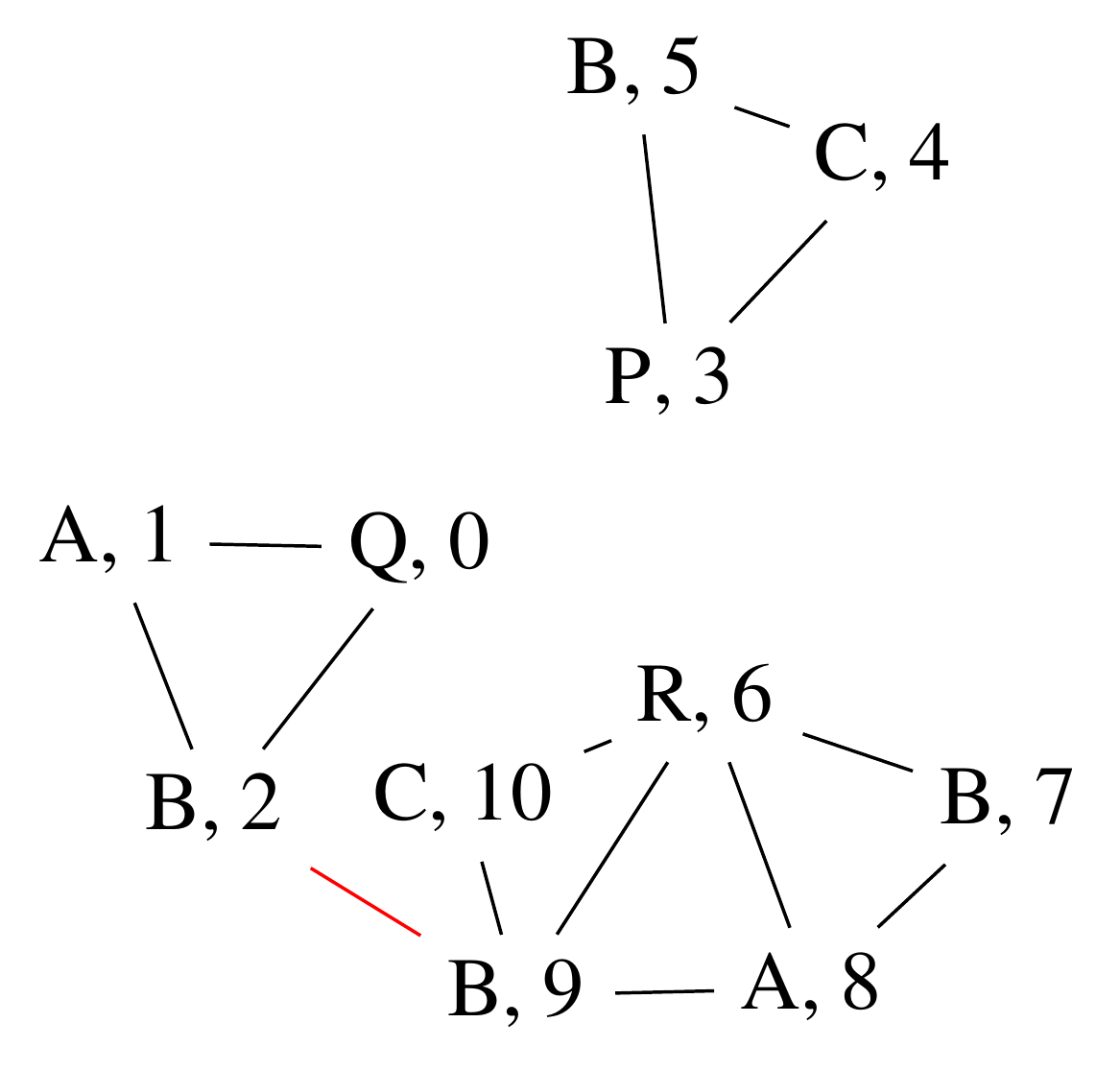}

\subsection*{Composition 6}
\begin{center}
\begin{tabular}{c|*{4}{c|}}
  \multicolumn{1}{c}{}
  & \multicolumn{1}{c}{$R_1^1$}
  & \multicolumn{1}{c}{$R_1^2$}
  & \multicolumn{1}{c}{$R_1^3$}	
  & \multicolumn{1}{c}{$R_1^\emptyset$}		\\ \cline{2-5}
  $L_2^1$	& 	&	& \textbullet$_1$ 	& 	\\ \cline{2-5}
  $L_2^2$	&	& 	& \textbullet	& 	\\ \cline{2-5}
\end{tabular}
\end{center}
\insertComp{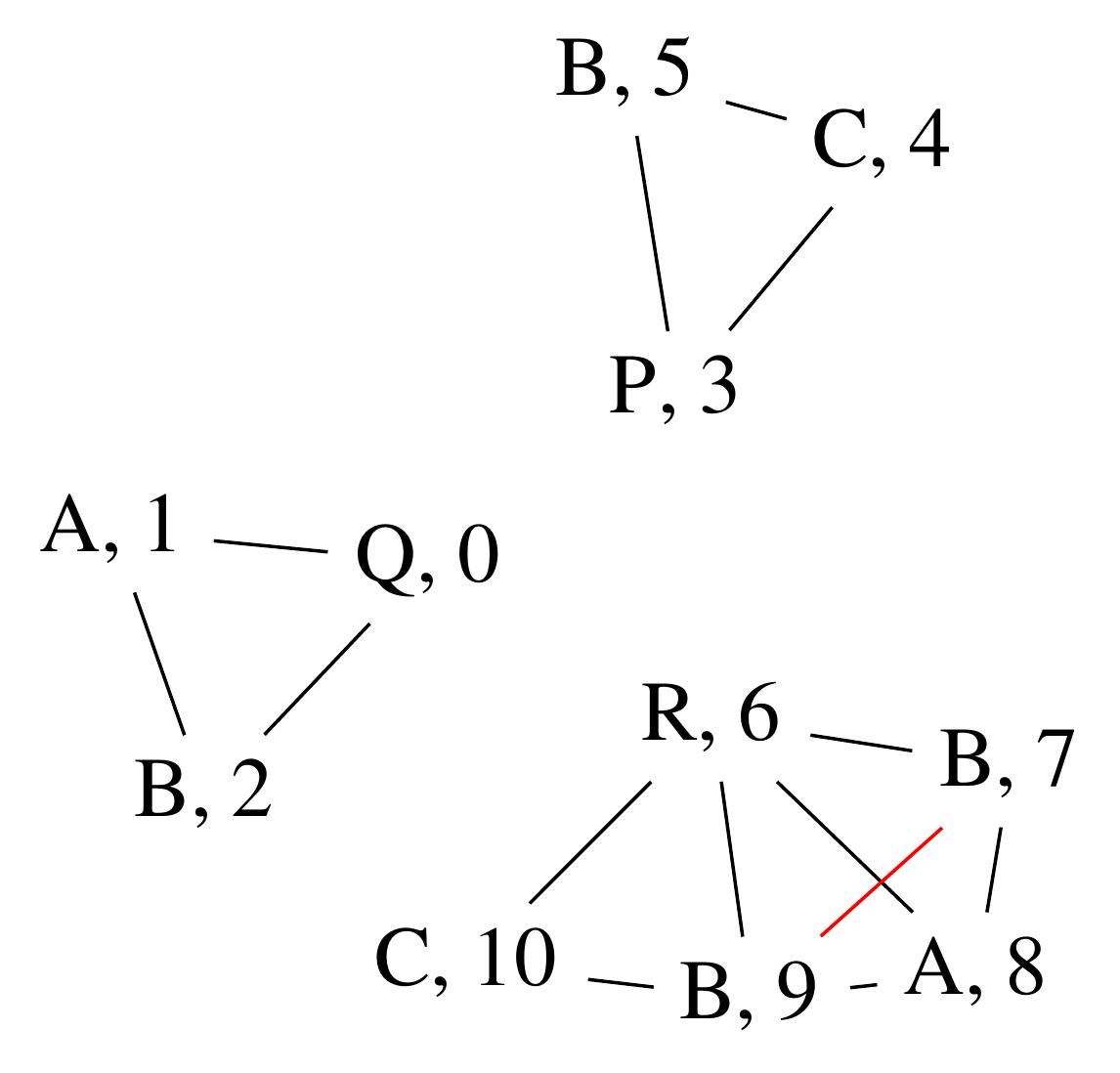}

\subsection*{Invalid Selection}
\begin{center}
\begin{tabular}{c|*{4}{c|}}
  \multicolumn{1}{c}{}
  & \multicolumn{1}{c}{$R_1^1$}
  & \multicolumn{1}{c}{$R_1^2$}
  & \multicolumn{1}{c}{$R_1^3$}	
  & \multicolumn{1}{c}{$R_1^\emptyset$}		\\ \cline{2-5}
  $L_2^1$	& 	&	& \textbullet$_2$ 	& 	\\ \cline{2-5}
  $L_2^2$	&	& 	& \textbullet	& 	\\ \cline{2-5}
\end{tabular}
\end{center}
This selection of submatches is invalid, as they are not disjoint (node $B,9$ would be matched twice).

\subsection*{Composition 7}
\begin{center}
\begin{tabular}{c|*{4}{c|}}
  \multicolumn{1}{c}{}
  & \multicolumn{1}{c}{$R_1^1$}
  & \multicolumn{1}{c}{$R_1^2$}
  & \multicolumn{1}{c}{$R_1^3$}	
  & \multicolumn{1}{c}{$R_1^\emptyset$}		\\ \cline{2-5}
  $L_2^1$	& 	&	& 	& \textbullet	\\ \cline{2-5}
  $L_2^2$	&	& 	& \textbullet	& 	\\ \cline{2-5}
\end{tabular}
\end{center}
\insertComp{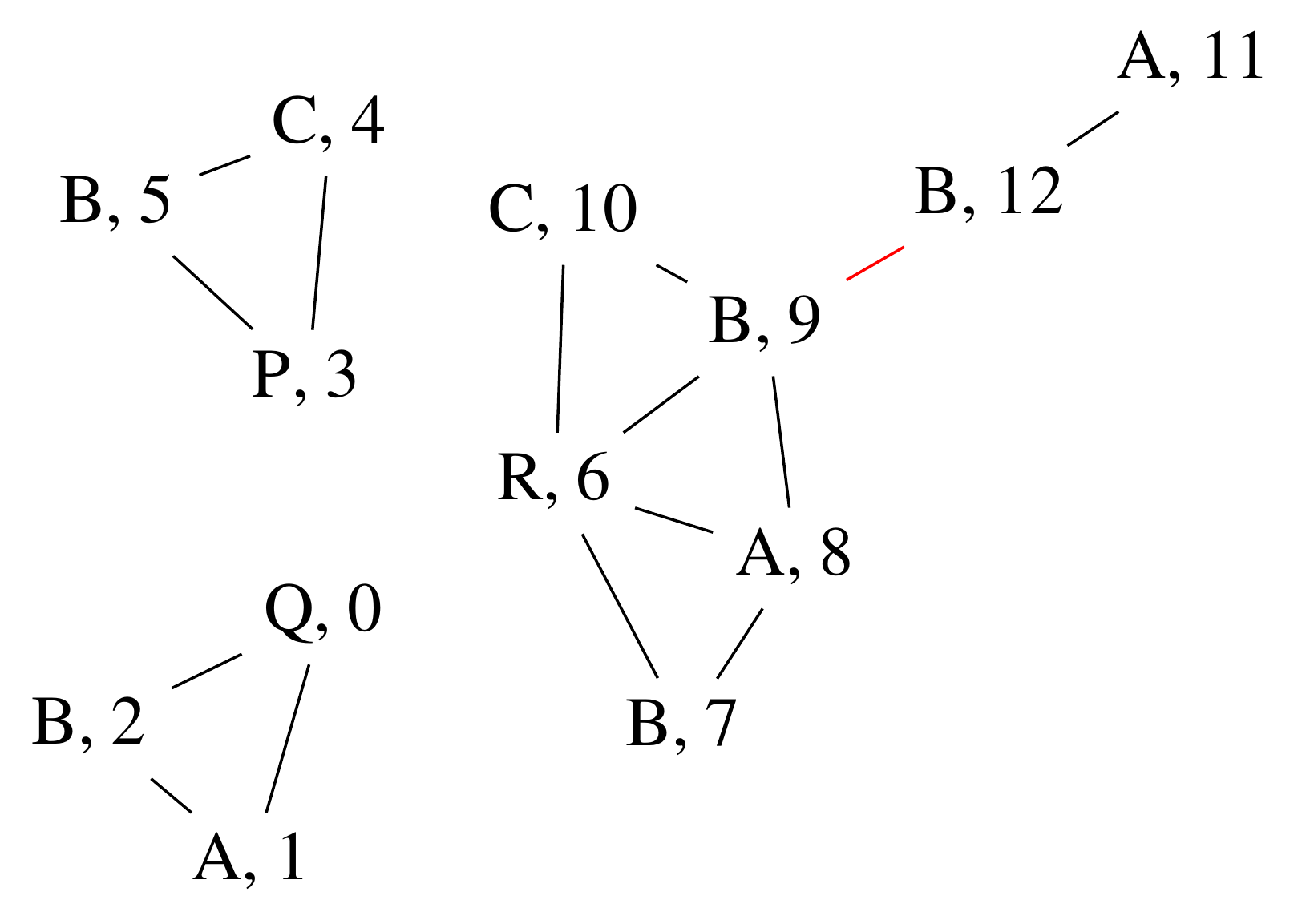}

\subsection*{Composition 8}
\begin{center}
\begin{tabular}{c|*{4}{c|}}
  \multicolumn{1}{c}{}
  & \multicolumn{1}{c}{$R_1^1$}
  & \multicolumn{1}{c}{$R_1^2$}
  & \multicolumn{1}{c}{$R_1^3$}	
  & \multicolumn{1}{c}{$R_1^\emptyset$}		\\ \cline{2-5}
  $L_2^1$	& \textbullet	&	& 	& 	\\ \cline{2-5}
  $L_2^2$	&	& 	& 	& \textbullet	\\ \cline{2-5}
\end{tabular}
\end{center}
\insertComp{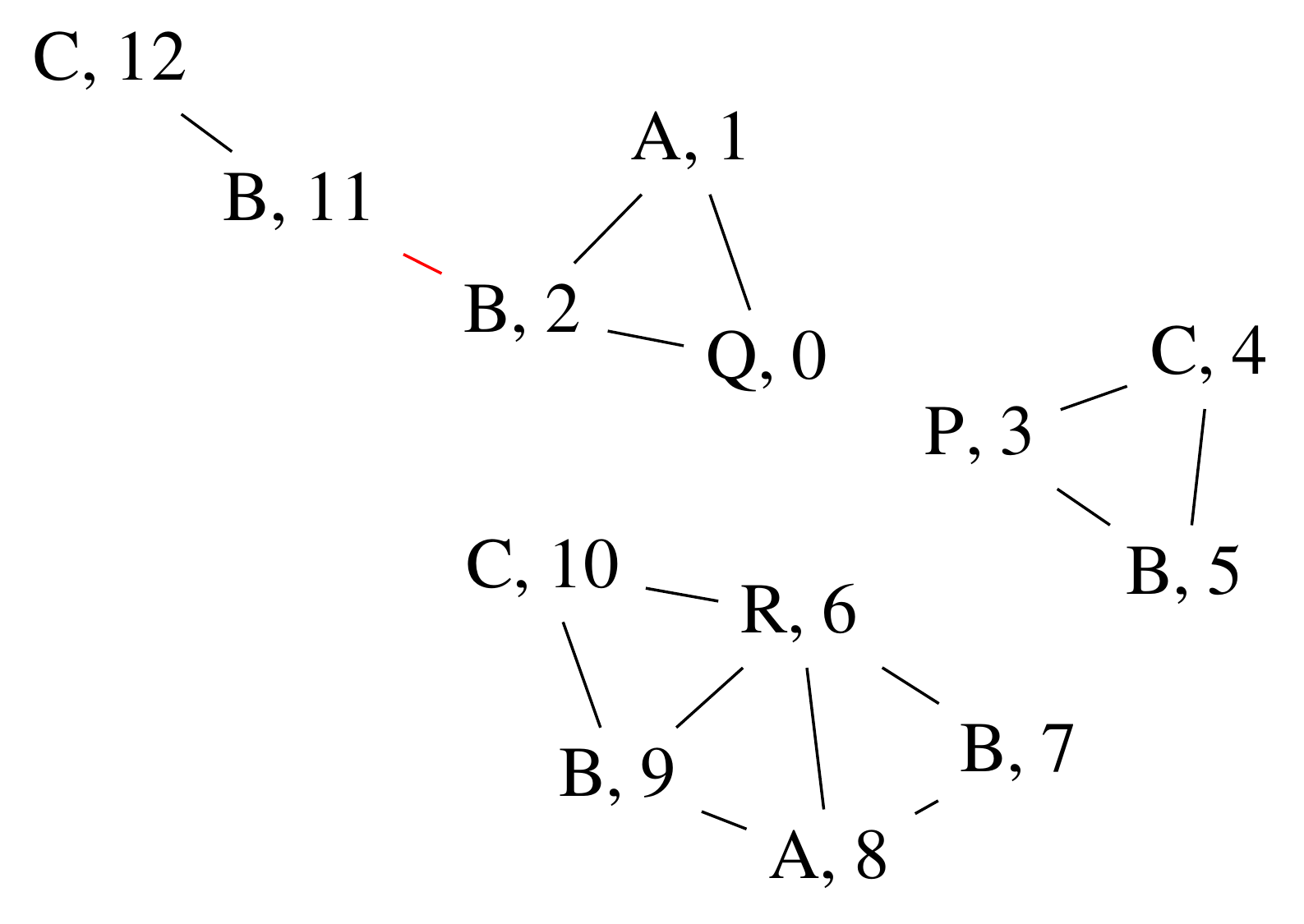}

\subsection*{Composition 9}
\begin{center}
\begin{tabular}{c|*{4}{c|}}
  \multicolumn{1}{c}{}
  & \multicolumn{1}{c}{$R_1^1$}
  & \multicolumn{1}{c}{$R_1^2$}
  & \multicolumn{1}{c}{$R_1^3$}	
  & \multicolumn{1}{c}{$R_1^\emptyset$}		\\ \cline{2-5}
  $L_2^1$	& 	&	& \textbullet$_1$	& 	\\ \cline{2-5}
  $L_2^2$	&	& 	& &\textbullet	 	\\ \cline{2-5}
\end{tabular}
\end{center}
\insertComp{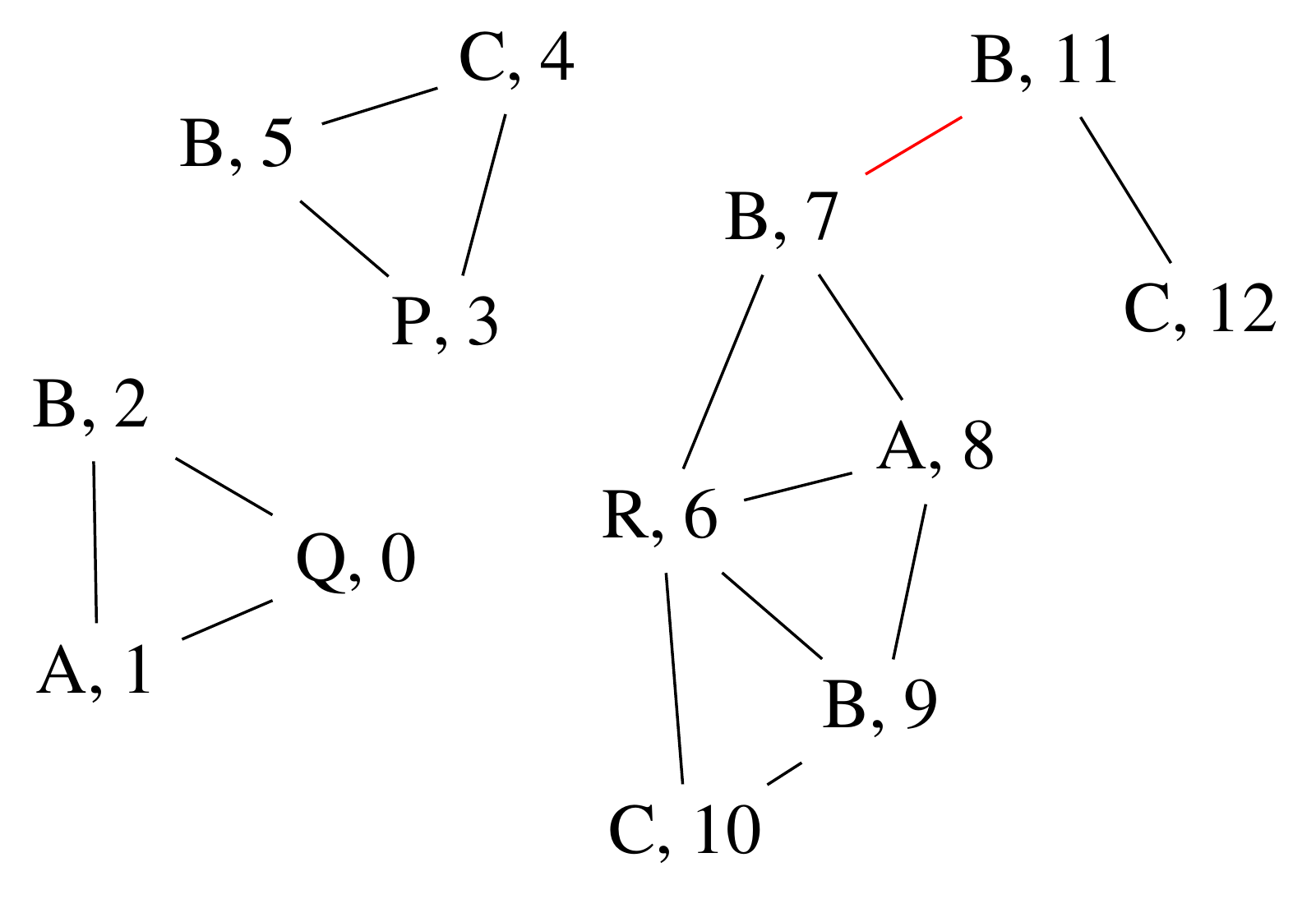}

\subsection*{Composition 10}
\begin{center}
\begin{tabular}{c|*{4}{c|}}
  \multicolumn{1}{c}{}
  & \multicolumn{1}{c}{$R_1^1$}
  & \multicolumn{1}{c}{$R_1^2$}
  & \multicolumn{1}{c}{$R_1^3$}	
  & \multicolumn{1}{c}{$R_1^\emptyset$}		\\ \cline{2-5}
  $L_2^1$	& 	&	& \textbullet$_2$	& 	\\ \cline{2-5}
  $L_2^2$	&	& 	& 	& \textbullet	\\ \cline{2-5}
\end{tabular}
\end{center}
\insertComp{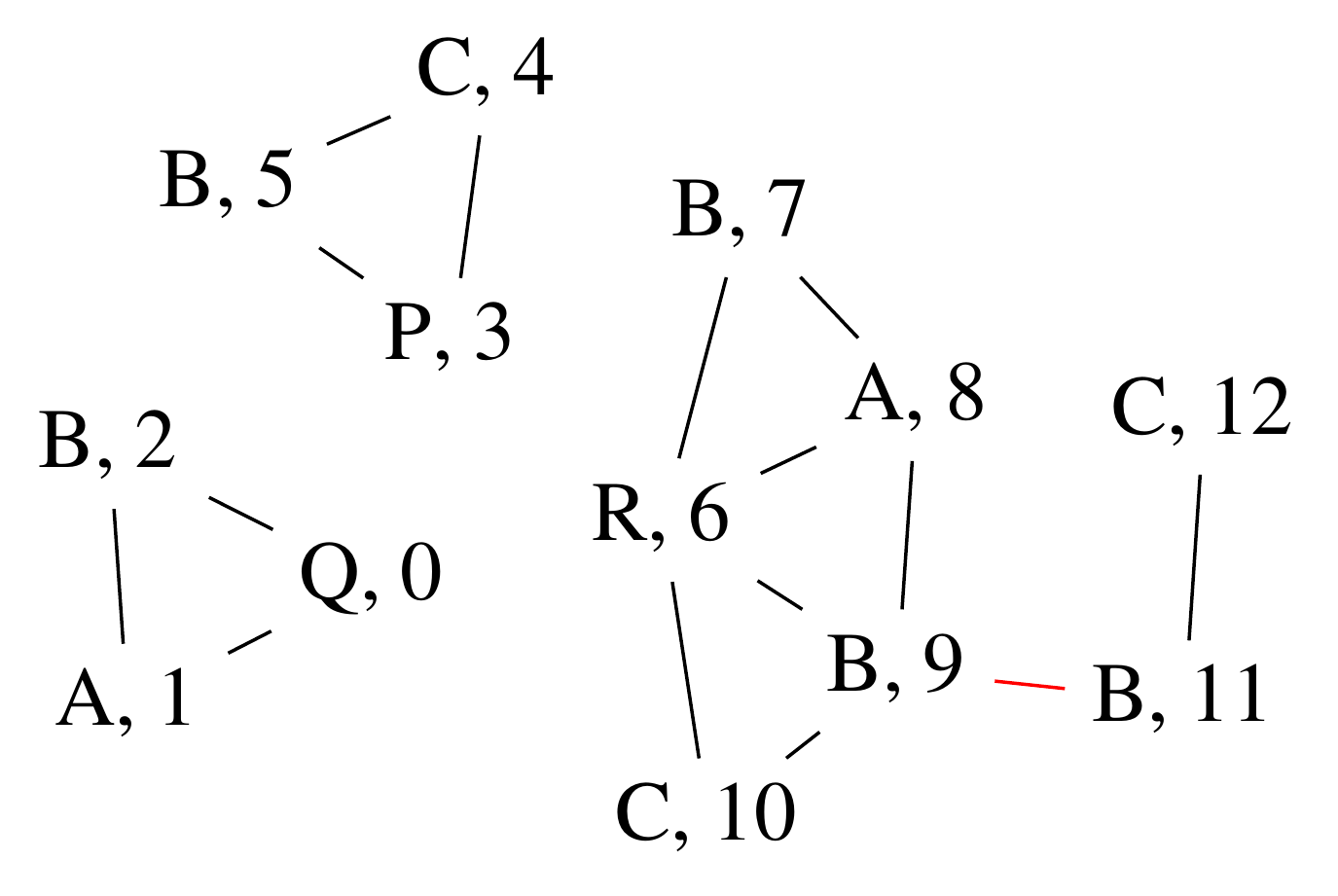}
\clearpage
\setcounter{page}{\value{endmaintextpage}}
\end{document}